\documentclass[letterpaper]{article} %
\usepackage[table]{xcolor}
\usepackage{aaai2026}  %
\usepackage{times}  %
\usepackage{helvet}  %
\usepackage{courier}  %
\usepackage[hyphens]{url}  %
\usepackage{graphicx} %
\usepackage{natbib}  %
\usepackage{caption} %
\usepackage{algorithm}
\usepackage[noend]{algorithmic}

\usepackage{newfloat}
\usepackage{listings}
\DeclareCaptionStyle{ruled}{labelfont=normalfont,labelsep=colon,strut=off} %
\floatstyle{ruled}
\newfloat{listing}{tb}{lst}{}
\floatname{listing}{Listing}
\usepackage{todonotes}
\usepackage{amsmath,amsthm, amssymb, amsfonts, mathtools}
\usepackage[capitalize,noabbrev]{cleveref}
\usepackage{subfigure}
\usepackage{stmaryrd}
\usepackage{xspace}

\usepackage{thmtools} 
\usepackage{thm-restate}

\theoremstyle{plain}

\theoremstyle{definition}

\theoremstyle{remark}

\usepackage{listings}

\usepackage{booktabs}
\newcommand{\acceptcell}{\cellcolor{green!20}A}
\newcommand{\rejectcell}{\cellcolor{red!20}R}

\lstdefinestyle{mystyle}{
    basicstyle=\ttfamily\small,
    breaklines=true,
    frame=single,
    columns=fullflexible,
    captionpos=b,
    language=Python,
    tabsize=2,
    keepspaces=true,
	numbers=none,
    literate={return}{{\bfseries return}}6
             {def}{{\bfseries def}}3
             {sample}{{\bfseries sample}}6,
    moredelim=**[is][\color{red}]{<<}{>>}, %
}

\newcommand{\size}[1]{\left| #1 \right|}

\newcommand{\E}{\mathop{\mathbb{E}}}

\newcommand{\remove}[1]{}

\newcommand{\Oh}{\mathcal{O}}

\newcommand{\tOh}{\widetilde{{\mathcal O}}}

\newcommand{\cX}{\mathcal{X}}

\newcommand{\eps}{\varepsilon}

\newcommand{\dtv}{\mathsf{d_{TV}}}

\newcommand{\uni}{\ensuremath{\mathsf{Unif}}\xspace}

\newcommand{\pois}{\mathsf{Poi}}
\newcommand{\Est}{\ensuremath{\mathsf{Est}}\xspace}
\newcommand{\tpa}{\ensuremath{\mathsf{TPA}}\xspace}
\newcommand{\Thresh}{\ensuremath{\mathsf{Thresh}}\xspace}
\newcommand{\Fail}{\ensuremath{\mathsf{Fail}}}
\newcommand{\Err}{\ensuremath{\mathsf{Error}}}

\newcommand{\bR}{\mathbb{R}}
\newcommand{\avgknown}{P\diamond Q}
\newcommand{\tilt}{\ensuremath{\mathsf{tilt}}}

\theoremstyle{plain}
\newtheorem{theo}{Theorem}
\newtheorem{lem}[theo]{Lemma}

\newtheorem{cl}[theo]{Claim}
\theoremstyle{definition}
\newtheorem{defi}[theo]{Definition}

\newcounter{mynotes}
\newcommand{\intcond}{\ensuremath{\mathsf{ICOND}}\xspace}
\newcommand{\cintcond}{\ensuremath{\intcond^\mathsf{Cont}}\xspace}
\newcommand{\tri}{\ensuremath{\mathsf{Tri}}\xspace}
\newcommand{\trih}{\ensuremath{\mathsf{Tri}}\xspace}
\newcommand{\Bad}{\ensuremath{\mathsf{Bad}}\xspace}
\newcommand{\Good}{\ensuremath{\mathsf{Good}}\xspace}
\newcommand{\accept}{\ensuremath{\mathsf{Accept}}\xspace}
\newcommand{\reject}{\ensuremath{\mathsf{Reject}}\xspace}

\newcommand{\unknown}{\ensuremath{P}\xspace}
\newcommand{\known}{\ensuremath{Q}\xspace}
\newcommand{\infident}{\ensuremath{\mathsf{ERtoltest}}\xspace}
\newcommand{\tvident}{\ensuremath{\mathsf{toltest}}\xspace}
\newcommand{\linf}{\ell_\infty}
\newcommand{\cunknown}[1]{\ensuremath{\ensuremath{P}^{#1}}\xspace}
\newcommand{\samples}{\ensuremath{\mathcal{X}}\xspace}
\newcommand{\dest}{\widehat{d}}
\newcommand{\pest}[1]{\widehat{P}_{#1}}
\newcommand{\rnd}[1]{\mathsf{rnd}(#1)}
\newcommand{\betacent}{\beta_{c}}
\newcommand{\Early}{\ensuremath{\mathsf{Early}}\xspace}
\newcommand{\Tiny}{\ensuremath{\mathsf{Tiny}}\xspace}
\newcommand{\contint}[1]{\llbracket#1\rrbracket}
\newcommand{\stp}{T}
\newcommand{\hhat}{\ensuremath{\mathsf{h}}\xspace}
\newcommand{\chat}{\ensuremath{\mathsf{H}}}
\newcommand{\samp}{\ensuremath{\mathsf{sampler}}\xspace}
\newcommand{\baseline}{\ensuremath{\mathsf{CubeProbe}}\xspace}
\newcommand{\tester}{\ensuremath{\mathsf{Lachesis}}\xspace}

\title{Instance Dependent Testing of Samplers using Interval Conditioning
\thanks{The full version of the paper, code, and benchmarks are available at https://github.com/uddaloksarkar/lachesis
.}
}
\author {
    Rishiraj Bhattacharyya\textsuperscript{\rm 1},
    Sourav Chakraborty\textsuperscript{\rm 2},
    Yash Pote\textsuperscript{\rm 3},
    Uddalok Sarkar\textsuperscript{\rm 2},
    Sayantan Sen\textsuperscript{\rm 4}
}
\affiliations {
    \textsuperscript{\rm 1}University of Birmingham\\
    \textsuperscript{\rm 2}Indian Statistical Institute Kolkata\\
    \textsuperscript{\rm 3}National University of Singapore\\
    \textsuperscript{\rm 4}Centre for Quantum Technologies,
 National University of Singapore
}
\date{}

\begin{document}
\maketitle

\begin{abstract}

Sampling algorithms play a pivotal role in probabilistic AI. However, verifying if a sampler program indeed samples from the claimed distribution is a notoriously hard problem.  Provably correct testers like Barbarik, Teq,
Flash, CubeProbe for testing of different kinds of samplers were proposed only in the last few years.
All these testers focus on the worst-case efficiency, and do not support verification of samplers over infinite domains, a case occurring frequently in Astronomy, Finance, Network Security, etc.

In this work, we design the first tester of samplers with instance-dependent efficiency, allowing us to test samplers over natural numbers. Our tests are developed via a novel distance estimation algorithm between an unknown and a known probability distribution using an \emph{interval conditioning} framework. The core technical contribution is a new connection with probability mass estimation of a continuous distribution. The practical gains are also substantial—our experiments establish up to 1000× speedup over state-of-the-art testers.

\end{abstract}

\section{Introduction}\label{sec:intro}
Sampling according to a distribution is ubiquitous in probabilistic AI. Its applications in efficient and accurate information extraction and decision making are crucial for advancements of data analytics and machine learning \cite{yuan2004simplifying,naveh2007constraint,mironov2006applications,morawiecki2013sat}. However, practical samplers are often based on heuristic techniques and lack formal guarantees, posing risks to transparency, reproducibility, security, and safety. As a result, verifying and certifying samplers efficiently has become a key challenge in building robust and trustworthy AI systems.

In the past seven years, several efficient testing algorithms have been developed~\cite{chakraborty2019testing,meel2020testing,PM21,pote2022scalable,pmlr-v206-banerjee23a,kumar2023tolerant} that are not only practically implementable but also come with formal correctness guarantees. These approaches build on algorithmic distribution testing~\cite{GoldreichR97,batu2001testing}, where sampler quality is assessed by measuring the distance between the target distribution and the actual distribution it produces. The central task is to determine whether an unknown distribution $\unknown$—accessed via sampling—is \emph{close} to or \emph{far} from a (potentially known) distribution $\known$. The efficiency of these tests depends on the number of samples drawn from $\unknown$.

Traditional distribution testing relies on i.i.d.\ samples from the unknown distribution~\cite{goldreich2017introduction,canonne2020survey,CanonneTopicsDT2022}. However, this approach often requires a number of samples polynomial in the support size, rendering it impractical for large domains. To overcome this limitation, the \emph{conditional sampling} model was introduced~\cite{CFGM16,CRS}, where an algorithm can query an oracle with a subset $S$ and receive samples from the conditional distribution $\unknown_{|S}$. This model is provably more powerful: for instance, it enables super-exponential reductions in sample complexity for testing distribution closeness. Despite its theoretical appeal, it was not immediately apparent how and in which contexts this more powerful sampling model could be implemented.

Recognizing the power of conditional sampling led to the development of several of its variants tailored to different contexts, typically by restricting the structure of the condition set 
$S$; for example, \emph{sub-cube conditions}~\cite{bhattacharyya2018property}, and \emph{pair} or \emph{interval conditions}~\cite{CRS}. These models enable practical implementations across diverse settings and have been used to design efficient verifiers for various samplers. For instance, standard conditional sampling has been applied to test CNF-samplers~\cite{chakraborty2019testing, meel2020testing, pmlr-v206-banerjee23a}, while sub-cube conditioning has enabled testers for self-reducible samplers~\cite{bhattacharyya2024testing, kumar2023tolerant}. Overall, conditional sampling has played a key role in enabling practical and efficient testing, while also motivating further study of sample complexity across different models.

In this paper, we focus on the \emph{interval conditional sampling} model, where distributions \( P \) and \( Q \) are defined over \( \mathbb{Z} \), and samples can be drawn from the conditional distribution \( P_{|S} \) for any interval \( S =  \contint{a,b} \) with \( a, b \in \mathbb{R} \). This model has been extensively studied, and the current state-of-the-art for estimating the \emph{total variation} distance between an unknown and a known distribution over domain \( [n] \) achieves a sample complexity of \( \tOh(\log^2 n) \)~\cite{bhattacharyya2024testing}, while the best lower bound is $\Omega(\log n/\log \log n)$~\cite{CRS}.

However, existing algorithms and their sample complexity bounds are largely oblivious to the finer structure of the distributions that are being tested. Most prior works focus on optimizing the \emph{worst-case} sample complexity—i.e., over all possible distributions on \( [n] \)—with complexity typically measured in terms of \(n \). This worst-case perspective has notable limitations. First, it does not extend to distributions over infinite discrete domains (e.g., \(  \mathbb{Z} \) or \( \mathbb{N} \)). Second, it misses opportunities for efficiency gains by ignoring the structural properties of the 
distributions \( \unknown \) and \( \known \).

In practice, many commonly encountered distributions—such as uniform, geometric, binomial, Poisson, or normal—exhibit smoothness properties (akin to Lipschitz continuity). This raises natural questions: \emph{Do we need to pay the worst-case sample complexity even when testing samplers over such smooth distributions? Can we design efficient verifiers for samplers over discrete, infinite domains? And, can the sample complexity be expressed in terms of a distribution's smoothness?}

We answer all these questions in the \emph{affirmative}.
We propose a new approach for estimating the total variation distance between the output of a sampler and a target distribution using interval conditional sampling. The sample complexity of our algorithm depends on a formal notion of smoothness of the distributions, allowing significantly improved performance when the known distribution is smooth. In particular, for a wide class of smooth distributions (including uniform, Gaussian, binomial, Poisson, and geometric), our algorithm achieves sample complexity significantly better than prior worst-case bounds of \( \tOh(\log^2 n) \). 

We introduce a novel algorithmic framework that imports techniques from the analysis of continuous distributions into the setting of discrete distribution testing under interval conditional sampling. This connection, previously unexplored, enables new algorithmic strategies for the discrete regime by systematically leveraging tools originally developed for continuous domains.

Our approach leads to several significant implications. First, it enables the design of an efficient tester for samplers over discrete infinite domains—an advancement we believe to be the first of its kind. Second, it introduces an instance-dependent framework for distance estimation in the interval conditional sampling model, where the sample complexity adapts to the structural properties of the distributions involved. This marks a shift from traditional worst-case analysis toward a more refined, distribution-aware perspective. Notably, this viewpoint aligns with the growing interest in instance-optimal algorithms, where the goal is to design algorithms whose performance matches the inherent difficulty of each individual instance~\cite{valiant2017automatic,hao2020data,feldmaninstance, diakonikolas2016new,DBLP:conf/focs/NarayananRTT24,DBLP:conf/focs/0001STTZ24}.

We provide an implementation of our algorithms and introduce a practical testing framework for a broad class of samplers known as \emph{inverse transform samplers}. Our results demonstrate how interval conditional sampling can be effectively applied in this context. The practical gains are substantial—our method achieves up to 1000× speedup over state-of-the-art worst-case algorithms~\cite{bhattacharyya2024testing} (see \Cref{fig:perf-exp}).

\paragraph{Organization}
The necessary preliminaries are in \Cref{sec:prelim}, followed by our contributions in \Cref{sec:ourresult}. The simulation of continuous interval conditioning is in \Cref{sec:simulate}, and our algorithms are described in \Cref{sec:algo}. Our experimental findings are in \Cref{sec:expresluts}. Due to space constraints, all proofs of theorems, lemmas, and claims are provided in the supplementary material.

\section{Preliminaries}\label{sec:prelim}

For a positive real number $a \in \bR^+$, $\rnd{a}$ denotes the nearest integer of $a$. $[n]$ denotes the set $\{1,2, \ldots, n\}$. For concise expressions and readability, we use the asymptotic complexity notion of $\widetilde{\Oh}(\cdot)$, where we hide polylogarithmic dependencies of the parameters. For integers $a, b \in [n]$, we use the standard bracket notation $[a, b]$ to denote the discrete interval $\{a, a+1, \ldots, b\}$, and for real numbers $u, v \in \mathbb{R}$, we denote the closed interval by $\contint{u,v}$. 
In this paper, we will deal with discrete distributions defined on a sample space $[n]$ (with $n \geq 2$) and continuous distributions defined on the real line $(-\infty, \infty)$.
The definition of discrete distributions naturally extends for $\mathbb{Z}$ such that $P(x) = 0$ for all $x \notin [n]$. 
For any $S \subseteq [n]$ or $S \subseteq \mathbb{R}$, we denote the probability mass of the set $S$ under a distribution $\unknown$ by $\unknown(S)$. For example, for integers $a\le b$ write $P([a,b])=\sum_{t=a}^b P(t)$. The Poisson distribution with parameter $\lambda$ is denoted by $\pois^\lambda$. We will use the Triangular distribution in this paper: Triangular distribution $\trih$ is defined for $x \in [-\tfrac12, \tfrac12]$ as
    \[
    \trih(x) = 
    \begin{cases}
        4x + 2 & -\frac{1}{2} \leq x \leq 0 \\
        -4x + 2 & 0 \leq x \leq \frac{1}{2}
    \end{cases}
    \]

We define the \emph{convolution} operation of a discrete distribution $\unknown$ with the continuous distribution $\trih$ as follows: 
\[\cunknown{\trih}(x) = \sum_{t \in [n]} \unknown(t) \trih(x - t) \]
Notably, $\cunknown{\trih}$ is a continuous distribution.

\begin{defi}[$\ell_{\infty}$ and $\dtv$-distance]
    For distributions $P$ and $Q$ over $[n]$ the multiplicative distance is defined as $$\ell_{\infty}(P,Q) = \max_{x \in [n]} \left| P(x)/Q(x)-1 \right|$$ The $\dtv$-distance between $P$ and $Q$ is defined as $$\dtv(P,Q) = \frac{1}{2}\sum_{x \in [n]}|P(x)-Q(x)|$$
\end{defi}

\begin{defi}[\tilt{}]
\label{defi:tilt}
Let $P$ be a distribution supported on $[n]$. Then for every $x \in [n]$ such that $P(x) > 0$, we define:
\[
\tilt_P(x)\;=\;\max\!\left\{\max_{y<x}\frac{P(y)}{P([y+1,x])}\;,\;\max_{y>x}\frac{P(y)}{P([x,y-1])}\right\},
\]
This captures the maximum ratio between the probability mass at any point $y$ and the total probability mass in the interval between $x$ and $y$ (excluding $y$).
\end{defi}

\paragraph{Access to Distributions}
To sample from a distribution $P$ is to draw an element $x \in [n]$ with probability $P(x)$. For a known distribution, the probability value $P(x)$ can be queried in constant time for any $x \in [n]$. In contrast, an unknown distribution allows only sample-based access, that is, drawing independent samples according to $P$. 
We also define sampling from a \emph{mixture distribution} of $P$ and $Q$, denoted by $\avgknown$ and defined as $\tfrac12 (P + Q)$, that is, drawing a sample from $P$ with probability $\tfrac{1}{2}$ and from $Q$ with probability $\tfrac{1}{2}$.
In this work, we assume a stronger form of access to the unknown distribution via an oracle, called the \emph{interval conditioning oracle}.

\begin{defi}[Interval Conditioning]
    \label{def:intcond}
    Given a distribution $\unknown$, the \intcond{} oracle (short for `interval conditioning') takes an interval $[i, j] \subseteq [n]$ and returns a sample drawn from $\unknown$ conditioned on $x \in [i, j]$, i.e., $x \sim \unknown_{|[i, j]}$. 
    We extend this to the continuous setting with the \cintcond{} oracle: given a distribution $\mu$ over $\mathbb{R}$ and an interval $\contint{u, v} \subseteq \mathbb{R}$, it returns a sample from $\mu$ conditioned on $x \in \contint{u, v}$, i.e., $x \sim \mu_{|\contint{u, v}}$.
\end{defi}

\subsection{Tootsie Pop Algorithm}

The Tootsie Pop Algorithm (\tpa)~\cite{banks2010using}, estimates the probability mass of a target interval within the domain of a continuous distribution. \tpa works by iteratively refining the interval until it reaches the target interval. 
We describe the Tootsie Pop Algorithm with a slight modification in \cref{alg:tpa}, adapted to operate using the \cintcond{} oracle. It takes as input a continuous distribution $\cunknown{\trih}$, an element $x$, number of iterations $r$, and parameters $\Thresh$ and $\delta$ which are related to the \intcond{} oracle. The algorithm returns an estimate of the probability mass of the interval $\contint{x - \tfrac12, x+ \tfrac12}$, or $\bot$ if it fails to estimate it within the given parameters. Within each iteration, the algorithm starts with the initial interval $\contint{-n,2n}$ (noting that $\cunknown{\trih}$ is supported only within $\contint{0, n}$) and refines to the target interval $\contint{x - \tfrac12, x + \tfrac12}$ by sampling from the distribution $\cunknown{\trih}$ conditioned on the current interval. It counts the number of steps ($\lambda$) taken to reach the target interval, and returns the average number of steps taken across all iterations. If the number of steps exceeds a threshold $\Thresh$ at any iteration or if the sampling fails, it returns $\bot$. 
\cite[Lemma 2]{banks2010using} has shown that the stopping time $\lambda$ follows a suitable Poisson distribution, which establishes the correctness of the algorithm \tpa{}, that is, $\lambda \sim \pois^\mu$, where $\mu = -\log \cunknown{\trih}(\contint{x - \tfrac12, x + \tfrac12})$.

\begin{algorithm}[tb]
\caption{$\tpa(\cunknown{\trih}, x, r, \Thresh, \delta, \theta)$}\label{alg:tpa}
\begin{algorithmic}[1]
    \STATE $k \gets 0$
    \FOR {$i = 1$ to $r$} \label{ln:est:for}
        \STATE $\lambda \gets 0$, $\beta \gets n$
        \WHILE { $\beta > \tfrac12$}
        \STATE $y \gets \cintcond\left(\cunknown{\trih}, x-\beta, x+\beta, \frac{\delta}{r \Thresh}, \theta\right)$
        \IF{$y = \bot$ or $\lambda\geq \Thresh$}
        \STATE {\bfseries Return} $\bot$
        \ENDIF

        \STATE $\lambda \gets \lambda + 1$, $\beta \gets |y-x|$
        \ENDWHILE
        \STATE $k \gets k + (\lambda-1)$\;
    \ENDFOR
    \STATE $\lambda \gets \frac{k}{r}$
    \STATE {\bfseries Return} $\lambda$
    \end{algorithmic}
\end{algorithm}

\subsection{Inverse Transform Samplers}
\label{sec:trs}

Inverse transform sampling is a powerful and widely used method for sampling from complex distributions. The core idea is to draw samples from a simpler distribution, typically uniform, and then apply a series of transformations to produce samples from the target distribution. This approach is especially effective when direct sampling from the target distribution is challenging—as in the case of Binomial~\cite{hormann1993generation}, Poisson~\cite{ormann1994transformed}, or Geometric distributions~\cite{fishman2001sampling}. In fact, standard libraries commonly implement samplers for these distributions using this paradigm.

When the inverse CDF of the target distribution is tractable (e.g., for the Geometric distribution; see \Cref{fig:geo_sampler_orig}), direct inverse transform sampling suffices. For distributions with intractable inverse CDFs, such as Binomial or Poisson, a common workaround involves using a \emph{hat distribution} \( \hhat \) that upper-bounds the target distribution \( \known \), i.e., there exists a constant \( C > 0 \) such that \( \known(x) \leq C \hhat(x) \) for all \( x \in \Omega \), and for which the CDF is easy to compute. Samples are then drawn from \( \hhat \) using:  
(1) inverse transform sampling, by drawing $u$ from uniform distribution over $\contint{0,1}$ and computing \( x = \chat^{-1}(u) \), followed by  
(2) rejection sampling, where \( x \) is accepted with probability proportional to \( \frac{\known(x)}{C\hhat(x)} \).

\begin{figure}[H]
\centering
\begin{minipage}{0.98\linewidth}
\begin{lstlisting}[
    language=Python,
    style=mystyle,
    caption={},
    literate={u_low}{{\textcolor{red}{u\_low}}}1
                {u_high}{{\textcolor{red}{u\_high}}}1
                {x_low}{{\textcolor{red}{x\_low}}}1
                {x_high}{{\textcolor{red}{x\_high}}}1
]
def Geometric(p):
    U = uniform(0,1) 
    return ceil(log(1 - U) / log(1 - p))
\end{lstlisting}
\end{minipage}
\caption{\small An inverse transform sampler for the geometric distribution.}
\label{fig:geo_sampler_orig}
\end{figure}

\section{Our Contributions}\label{sec:ourresult}

Our main contributions are two algorithms $\tvident$ and $\infident$, the first set of algorithms for identity testing of distributions in the interval conditioning model with instance-dependent bounds. We also developed the first practical realization of the interval conditioning oracle to test inverse transform samplers. Our technical contributions are summarized in the following theorem.

\begin{restatable}{theo}{maintvident}\label{thm:tvident}
    Let $\unknown$ be an unknown distribution and $\known$ a known distribution over $\mathbb{Z}$. 
    Given access to $\intcond(\unknown)$, accuracy parameters $\eps, \eta \in (0,1)$ with $\eta > \eps$, and confidence parameter $\delta \in (0,1)$, the algorithms $\tvident$ and $\infident$ satisfy the following:
  \begin{itemize}
      \item $\tvident$ can distinguish between the cases $\dtv(\unknown, \known) \leq \eps$ and $\dtv(\unknown,\known) \geq \eta$ with probability $\geq 1- \delta$.

     The expected number of $\intcond$ queries is
    \[
      \tOh\left(\frac{1}{(\eta - \eps)^4} \E_{x \sim \avgknown} \left[\tilt_{\avgknown}(x) \cdot \log^2 \frac{1}{\avgknown(x)} \right] \right).
    \]
 where $\avgknown$ is the distribution defined by $\avgknown(x)=(P(x)+Q(x))/2$.  
 \item 
 $\infident$ can distinguish between the cases $\linf(\unknown, \known)\leq 2\eps$ and $\dtv(\unknown,\known) \geq \eta$ with probability $\geq 1- \delta$. The expected number of oracle queries made by the $\infident$ is at most
\[\tOh\left(\frac{1}{(\eta - \eps)^4} \E_{x\sim \unknown}\left [ \tilt_\known(x) \cdot  K(x)\right]\right)\]
Where $K(x) = \min\left(\log^2 \frac{1}{\unknown(x)}, \log^2 \frac{1}{\known(x)}\right)$.
 \end{itemize} 
\end{restatable}

Both our algorithms use the Tootsie Pop Algorithm (\tpa). However, \tpa crucially needs \intcond{} to a continuous distribution, whereas the distribution $\unknown$ under test is a discrete distribution. To address this, we
consider the convolution of $\unknown$ with a continuous distribution. We needed a distribution to convolve with such that we can simulate drawing samples (or interval conditional samples) from the convolved distribution using \intcond{} on $\unknown$. For this, we chose the triangular distribution and work with the distribution $\cunknown{\trih}$. Finally,  we employ rejection sampling in (\Cref{alg:intcond}) to simulate \cintcond{} on $\cunknown{\trih}$ using \intcond{} on $\unknown$. This simulation is presented in Section~\ref{sec:simulate}.

The crucial subroutine that we need for both our algorithms $\tvident$ and $\infident$ is $\Est$, which is used to estimate the value of $\unknown(x)$ for any $x\in \mathbb{Z}$. This subroutine uses the \tpa using the \cintcond{} oracles to $\cunknown{\trih}$. The expected runtime of $\Est$ depends on the smoothness (\tilt) of $\unknown$, and this is where we observe that our algorithms are instance-dependent, that is, the runtime varies with the quality of the unknown distribution $\unknown$. Finally, using $\Est$, we describe our algorithms. The algorithms $\Est$,  $\tvident$ and $\infident$ are described in Section~\ref{sec:algo}. 

\paragraph{Experimental Results} 
We extended our algorithms to build the first practical tester for the broader class of samplers—namely, inverse transform samplers. The key technical challenge in this direction is implementing the interval conditioning oracle for such samplers. 
In this work, we designed the first concrete implementation of such an \intcond{} oracle, specifically tailored to operate with inverse transform samplers. This implementation is non-trivial and involves formal guarantees to ensure correctness. The design and methodology of this oracle are described in detail in \cref{sec:expresluts}.
We integrate this oracle to construct our practical tester \tester{}. Empirical results show that \tester{}  significantly outperforms existing methods for testing samplers, offering substantial improvements in efficiency.

\paragraph{Technical Novelty}
\textsc{From Continuous to Discrete Testing:}
The core of our technique is in efficient probability mass estimation exploiting the \emph{structure} of the distributions. The main technical novelty is in adapting the Tootsie Pop Algorithm (\tpa), which is used to approximate the integral of nonnegative functions over high-dimensional continuous spaces \cite{huber2010TPA}. We crucially observe that 
\tpa could be employed to estimate the probability mass of a subset of the domain if one could establish interval conditioning over the reals using the oracle that allows interval conditioning samples over $\mathbb{Z}$. To achieve this,  we innovate a technique for achieving interval conditioning over reals via extended distribution and rejection sampling. 
The efficiency of this transformation depends on the ``smoothness'' of the distributions.

\paragraph{Notable Features of our algorithms}
There are a couple of notable and novel features of our algorithms that we would like to highlight:

\ 

\noindent\textsc{Testing Distributions over infinite support:}
The sample complexities of our algorithms are independent of the domain size. This allows our algorithms to test distributions over infinite support, e.g., $\mathbb{Z}$, often with very few queries, for a large class of distributions, for example, \emph{log-concave distributions}~\cite{bagnoli2005log, lovasz2007geometry}. For example, consider the case when $\known$ is a Poisson distribution with parameter $\lambda >0$. In such a scenario, our algorithm $\infident$ can efficiently distinguish whether an unknown distribution $\unknown$ is close to or far from $\known$, with an expected number of queries $\tOh\left(\log^2\lambda\right)$.

\ 

\noindent\textsc{Testing Distributions over Small support:} 
Many times, the sampler's output is concentrated over a small set, following the target distribution perfectly over that set. Consider, for example, the case when $\unknown$ is \emph{uniform} over a subset $S \subset [n]$, such that, $|S| \ll n$. The best known algorithm, \baseline~\cite{bhattacharyya2024testing}, requires $\tOh\left(\frac{\log^2 n}{(\eta - \eps)^4}\right)$ queries to the interval conditioning oracle to distinguish between $\linf(\unknown, U)\leq 2\eps$ and $\dtv(\unknown,U) \geq \eta$, where $U$ denotes the uniform distribution over $[n]$. The following corollary of \Cref{thm:tvident} provides a better result.

\begin{restatable}{coro}{tolerantuniformity}\label{coro:tolerantuniformity}    
Let $\unknown$ be an unknown distribution, promised to be supported on a subset $S \subseteq [n]$, and let $U$ denote the uniform distribution over $[n]$. Our algorithm $\infident$ can distinguish between $\linf(\unknown, U)\leq 2\eps$ and $\dtv(\unknown,U) \geq \eta$, with at most $\tOh\left(\frac{\log^2 |S|}{(\eta - \eps)^4}\right)$ queries to the interval conditioning oracle in expectation.
\end{restatable}

\section{Simulating \cintcond{} from \intcond{}}
\label{sec:simulate}

We will use the Tootsie Pop algorithm (\tpa, \Cref{alg:tpa}), which works in the context of continuous distributions. However, we only have access to the \intcond{} oracle for an unknown discrete distribution $\unknown$. In this section, we will devise a method of simulating conditional samples from a continuous distribution with \intcond{} query access to $\unknown$, in particular, we will introduce an algorithm that simulates the \cintcond{} oracle for a continuous distribution $\cunknown{\trih}$.

Throughout this section, we assume access to a sampling procedure for the triangular distribution $\trih$, an assumption justified by well-established sampling methods for triangular distributions~\cite{kotz2004beyond}. 
In \cref{alg:intcond}, we describe the construction of the continuous interval conditioning oracle \cintcond{}, using the basic \intcond{} oracle and sampling access to $\trih$, implemented via a rejection sampling–based approach.

Given $\intcond$ access to the unknown distribution $\unknown$, interval $\contint{u, v}$, the confidence parameter $\delta$ and a stopping parameter $\theta$, \cintcond first set the maximum number of attempts to $T = (2\theta + 1)\log \tfrac1\delta$. Then it obtains a sample $x$ from \intcond oracle conditioned on the interval $[\rnd{u}, \rnd{v}]$. Next, it also obtains another sample $r$ independently from the triangular distribution $\trih$. If $(x+r)$ belongs to the interval $\contint{u, v}$, \cintcond outputs $(x+r)$ as a sample from $\cunknown{\trih}$. Otherwise, it restarts the same process. If no valid sample is obtained in $T$ iterations, \cintcond outputs $\bot$ and terminates the algorithm. The following theorem states the correctness of the \cref{alg:intcond}.

\begin{algorithm}[t]
    \caption{$\cintcond(\cunknown{\trih}, u, v, \delta, \theta)$}
    \label{alg:intcond}
    \begin{algorithmic}[1]

    \STATE $T \gets (2\theta + 1) \log \frac{1}{\delta}$
    \FOR{$i \in \{0 \ldots \stp\}$}
        \STATE $x \gets \intcond(\unknown, [\rnd{u}, \rnd{v}])$ \label{ln:intcond:sample}
        \STATE $r \gets \trih$
        \IF{$(x + r) \in \contint{u, v}$}
            \RETURN $x + r$ \label{alg2:valreturn}
        \ENDIF
    \ENDFOR
    \RETURN $\bot$
    \end{algorithmic}
\end{algorithm}

\begin{theo}[Correctness of \cintcond{}]\label{thm:simulation}
    Let $\unknown$ be a discrete distribution defined over $[n]$ and  $u, v$ are any real numbers from $\contint{0, \ n}$ with  $v \geq u + 1$. 
    Then the algorithm $\cintcond$ simulates the \cintcond{} oracle for the conditional continuous distribution $\cunknown{\trih}_{|\contint{u,v}}$. Moreover, if the algorithm terminates with a valid real number in $\contint{u,v}$, it makes at most $\Oh(L)$ calls to the \intcond{} oracle in expectation, where $L = \max\left\{\tfrac{\unknown(u)}{\unknown([u+1, v])}, \tfrac{\unknown(v)}{\unknown([u, v-1])}\right\}$.
\end{theo}

We now outline the two key components underlying the proof of \cref{thm:simulation}, deferred to the supplementary material. 
First, we argue that when the algorithm $\intcond$ does output a real value (i.e., the output is not $\bot$), then the value returned is a real number from $\contint{u,v}$ drawn according to the distribution $\cunknown{\trih}_{|\contint{u,v}}$. 
The main observation here is that the number $(x + r)$ is a real number between $\contint{\rnd{u} - \tfrac{1}{2}, \rnd{v} + \tfrac{1}{2}}$ drawn according to the distribution $\cunknown{\trih}_{|\contint{\rnd{u} - \tfrac{1}{2}, \rnd{v} + \tfrac{1}{2}}}$. If $(x+r)$ is not within the range $\contint{u,v}$, then the process is repeated for $T$ times or until a number between $\contint{u,v}$ is picked, whichever occurs first.  

Second, we show that if $\theta$ is large enough, the probability that the algorithm $\intcond$ outputs $\bot$ is small. Informally speaking, while calling the \intcond{} oracle on the interval $[\rnd{u}, \rnd{v}]$, as long as $\theta \geq L$, where $L = \max\left\{\tfrac{\unknown(u)}{\unknown([u+1, v])}, \tfrac{\unknown(v)}{\unknown([u, v-1])}\right\}$, the chance of $(x+r)$ not falling in the range $\contint{u,v}$ in any of the $T$ rounds is very small. And if the algorithm does not output $\bot$, it takes at most $\Oh(L)$ calls to the \intcond{} oracle in expectation. 

\section{Our Algorithms}\label{sec:algo}
We now present our algorithms, as stated in \Cref{thm:tvident}, assuming access to \intcond queries on the unknown distribution~$\unknown$. We begin by 
describing a crucial subroutine, \Est,  that estimates the value of $\unknown(x)$ using the \tpa algorithm. Then we will describe our algorithms $\tvident$ and $\infident$, both of which use $\Est$ as a subroutine.

\subsection{Probability Mass Estimator: \Est}
Given access to the $\intcond$ oracle for the unknown distribution $\unknown$, along with parameters $\zeta$, $\delta$, $B$, and a target element $x \in [n]$, \Est estimates $\unknown(x)$ in two phases using the \tpa algorithm.
In the first phase, it calls \tpa to obtain a coarse estimate $\lambda$. If this call fails (returns $\bot$), \Est also returns $\bot$. Otherwise, it computes a refined iteration bound $r_2$ based on $\lambda$, and makes a second \tpa call. If the second call fails, \Est again returns $\bot$; otherwise, it returns $e^{-\lambda}$ as the estimate of $\unknown(x)$.
To limit the number of iterations in \tpa, \Est uses thresholds $\Thresh_1$ and $\Thresh_2$, determined by $B$, in each phase.
Formally, the correctness of \Est is stated in the following lemma.

\begin{restatable}[Correctness of \Est]{lem}{est}\label{lem:est}
    Given \intcond access to $\unknown$, an element $x \in [n]$, and parameters $\zeta, \delta \in (0,1)$, the algorithm \Est returns a value $\widehat{\unknown}_x$. If $\unknown(x) \geq \frac{1}{\theta}$, then with probability at least $1 - \delta$, $\widehat{\unknown}_x \in (1 \pm \zeta)\cdot \unknown(x)$ holds. The expected number of \intcond queries performed by \Est is $\tOh\left(\frac{1}{(\eta - \eps)^2} \E_{x\sim \unknown}\left[\min(\tilt_\unknown(x), \theta) \cdot \log^2 \tfrac{1}{\unknown(x)}\right]\right)$.
\end{restatable}

\begin{algorithm}[tb]
    \caption{$\Est(\unknown, x, \zeta, \delta, B, \theta)$}
    \label{alg:est}
    \begin{algorithmic}[1]
        \STATE \textbf{$\blacktriangleright$ First phase of \tpa for early estimate}
        \STATE $r_1 \gets 2 \log \frac{8}{\delta}$
        \STATE 
        $\Thresh_1 \gets B + \log\frac{2r_1}{\delta} + \sqrt{\log^2 \frac{2r_1}{\delta} + 2B \log \frac{2r_1}{\delta}}$
        \STATE  $\lambda \gets \tpa(\cunknown{\trih}, x, r_1, \Thresh_1, \frac{\delta}{4}, \theta)$
        \IF{$\lambda = \bot$} \RETURN $\bot$ \ENDIF
        \STATE \textbf{$\blacktriangleright$ Second phase of \tpa to refine estimate}
        \STATE New iterations:
        $r_2 \gets \frac{2(\lambda + \sqrt{\lambda} + 2 + \log(1+\zeta))}{\log^2(1+\zeta)} \cdot \log \frac{16}{\delta}$
        \STATE Compute refined threshold:
        $\Thresh_2 \gets B + \log\frac{2r_2}{\delta} + \sqrt{\log^2 \frac{2r_2}{\delta} + 2B \log \frac{2r_2}{\delta}}$
        \STATE $\lambda \gets \tpa(\cunknown{\trih}, x, r_2, \Thresh_2, \frac{\delta}{4}, \theta)$
        \IF{$\lambda = \bot$} \RETURN $\bot$ \ENDIF
        \RETURN Final estimate: $e^{-\lambda}$
    \end{algorithmic}
\end{algorithm}

\subsection{Our Distance Estimator: \tvident}\label{sec:indtvtv}
{
We now describe our algorithm,  \tvident.  Given \intcond query access to an unknown distribution \unknown, a known distribution \known, and parameters $\eps, \eta$ with $\eta > \eps$, outputs \accept if $\dtv(\unknown, \known)\leq \eps$ and outputs \reject if $\dtv(\unknown, \known)\geq \eta$. 
We will first describe the algorithm and then prove its correctness. The algorithm \tvident (\Cref{alg:indl1test}) uses $\Est$ (\Cref{alg:est}) as subroutine. 
The algorithm works by sampling from a mixture distribution $\avgknown := \frac{\unknown + \known}{2}$.

\tvident first obtains a multiset $\samples$ of $t'$ samples from $\avgknown$. For each sample $x_i \in \samples$, it calls the subroutine \Est to estimate the probability mass of $P(x_i)$ with a maximum threshold $B$ and confidence parameter $\frac{\delta}{4t'}$. If \Est returns $\bot$ for any $x_i \in \samples$, \tvident discards the sample from $\samples$. If the number of samples in $\samples$ is less than $t$, it outputs \reject and terminates the algorithm. Otherwise, it computes $\dest$ depending on the estimates obtained. 
Finally, if the estimate $\dest$ is more than $\frac{\eta+ \eps}{4}$, it outputs \reject and terminates the algorithm. Otherwise it outputs \accept, and declares that $\unknown$ and $\known$ are $\eps$-close in $\dtv$ distance. 
The threshold $B$ (computed based on the min-entropy of $\known$) dictates the sample complexity of $\Est$.

\begin{algorithm}[t]
    \caption{$\tvident(\unknown, \known, \eps, \eta, \delta)$}
    \label{alg:indl1test}
    \begin{algorithmic}[1]
    \STATE $\eps' \gets \tfrac\eps2$, $\eta' \gets \tfrac\eta2$ 
    
    \STATE $\zeta \gets \frac{\eta' - \eps'}{\eta' - \eps' + 2}$, $t \gets \frac{8}{(\eta' - \eps')^2} \log \frac{4}{\delta}$
    \STATE $k \gets 1 + \frac{1}{t} \log \frac{4}{\delta} + \sqrt{\frac{1}{t} \log^2 \frac{4}{\delta} + \frac{2}{t} \log \frac{4}{\delta}}$
    \STATE $t' \gets 3 k t$
    \STATE $\avgknown=\frac{P+Q}{2}$
    \STATE Draw a multiset $\samples = \{x_1, \ldots, x_{t'}\}$ from $\avgknown$
    \STATE $B \gets \log \theta  + \log(1 + \eps') $, \ $\theta \gets \frac{1}{\known_{\min}}$
    
    \FOR{$i = 1$ to $t'$}
        \STATE $\widehat{\unknown}_{x_i} \gets \Est(\avgknown, x_i, \eps', \zeta, \frac{\delta}{4 t'}, B, \theta)$
        \IF{$\widehat{\unknown}_{x_i} = \bot$}
            \STATE $\samples.\mathsf{remove}(x_i)$
        \ENDIF
    \ENDFOR
    \IF{$|\samples| < t$}
        \RETURN \reject
    \ENDIF
    
    \STATE $\dest \gets \frac{1}{|\samples|} \sum_{x \in \samples} \max\left(0, 1 - \frac{\known(x)}{\widehat{\unknown}_x}\right)$
    
    \IF{$\dest > \frac{\eta + \eps}{4}$}
        \RETURN \reject
    \ENDIF
    
    \RETURN \accept
    \end{algorithmic}
\end{algorithm}

\subsection{The Early Reject Tester: \infident}\label{sec:indlinftv}

Now we modify the identity tester \tvident to obtain a simpler early reject identity tester $\infident$ (\Cref{alg:indlinftest}) that outputs \accept if $\linf(\unknown, \known)\leq \eps$ and outputs \reject if $\dtv(\unknown, \known)\geq \eta$. 
The primary difference between \infident and \tvident is that \infident uses the full power of the \Est subroutine that also uses a threshold parameter $B$ to control the running time of the \tpa subroutine. While \tvident uses a fixed threshold $B$ that is independent of the sample $x_i$, \infident uses a sample dependent threshold $B$ that is computed based on the probability of $x_i$ in the known distribution $\known$. This allows \infident to run faster than \tvident, as it can adaptively adjust the threshold based on the sample being processed. Also, if \Est returns $\bot$ for any sample $x_i$, \infident immediately outputs \reject and terminates.

\begin{algorithm}[t]
    \caption{$\infident(\unknown, \known, \eps, \eta, \delta)$}
    \label{alg:indlinftest}
    \begin{algorithmic}[1]

    \STATE $\zeta \gets \frac{\eta - \eps}{\eta - \eps + 2}$,  $t \gets \frac{8}{(\eta - \eps)^2} \log \frac{4}{\delta}$
    
    \STATE Draw a multiset $\samples = \{x_1, \ldots, x_t\}$ from $\unknown$
    
    \FOR{$i = 1$ to $t$}
        \IF{$\known(x_i) = 0$}
            \RETURN \reject
        \ENDIF
        \STATE $B \gets \log\frac{1+2\eps}{\known(x_i)}$, \ $\theta \gets \frac{1+\eps}{1-\eps}\tilt_\known(x_i)$
        \STATE $\pest{x} \gets \Est(\unknown, x_i, \zeta, \frac{\delta}{4t}, B, \theta)$
        \IF{$\pest{x} = \bot$}
            \RETURN \reject
        \ENDIF
    \ENDFOR
    
    \STATE $\dest \gets \frac{1}{t} \sum_{i=1}^t \max\left(0, 1 - \frac{\known(x_i)}{\pest{x_i}}\right)$
    
    \IF{$\dest > \frac{\eta + \eps}{2}$}
        \RETURN \reject
    \ENDIF
    
    \RETURN \accept
    \end{algorithmic}
\end{algorithm}

\section{Testing Inverse Transform Samplers}\label{sec:expresluts}

In this section, we extend \infident{} and \tvident{} to design the tester \tester{} for inverse transform samplers. The tester \tester{} has two modes: (1) \infident{} mode, and (2) \tvident{} mode. We use \tester{} to verify the correctness of three common samplers implemented following NumPy's design standards: (1) Geometric distribution sampler~\cite{fishman2001sampling}, (2) a Binomial sampler \cite{hormann1993generation}, (3) a Poisson sampler \cite{ormann1994transformed}. 
We assume full access to the program code, allowing us to inspect and modify the implementation. We note that despite having complete visibility, it is \#P-Hard to explicitly determine the exact distribution of a randomized program's output~\cite{holtzen2020scaling}. Thus, a tester has to employ a sampling based technique, possibly using conditional samples, which approximates the probability mass function of the program's output distribution.
We first describe how \intcond{} can be implemented for programs that realize the inverse transform samplers.

\subsection{\intcond{} Implementation}
\label{sec:icondimp}
In inverse transform samplers, the inverse sampling component is responsible for the primary sample generation. Therefore, the \intcond{} oracle for such samplers can be implemented by conditioning the output of this inverse step to fall within a specified interval. This, in turn, reduces to sampling uniformly from the corresponding pre-image of that interval under the transformation. \Cref{fig:geo_sampler} illustrates the implementation of \intcond{} for the geometric distribution sampler, a simple example of the inverse transform sampler.
\begin{restatable}[\intcond{} Implementation]{cl}{expcl}
    \label{thm:exp}
    Let $\samp$ be an inverse transform sampler that uses a hat distribution $\hhat$ with cumulative distribution function $\chat$. Then, the interval conditioning query $\intcond(\samp, [a, b])$ is equivalent to $\intcond(\uni, \contint{\chat(a), \chat(b)})$, where $\uni$ denotes the uniform distribution over $\contint{0,1}$.
\end{restatable}    
    
\begin{table*}[ht]
    \centering

\begin{tabular}{lcccccccccccc} 
\toprule
Parameters & \multicolumn{2}{c}{1} & \multicolumn{2}{c}{2} & \multicolumn{2}{c}{3} & \multicolumn{2}{c}{4} & \multicolumn{2}{c}{5} & \multicolumn{2}{c}{6} \\
(n, p) or $\mu$ & Dec. & \# Calls & Dec. & \# Calls & Dec. & \# Calls & Dec. & \# Calls & Dec. & \# Calls & Dec. & \# Calls \\
\midrule
31306, 0.16 & \acceptcell & 46K & \rejectcell & 38K & \rejectcell & 9830K & \rejectcell & 9874K & \rejectcell & 9874K & \rejectcell & 9874K \\
83836, 0.42 & \acceptcell & 49K & \rejectcell & 41K & \rejectcell & 7124K & \rejectcell & 6966K & \rejectcell & 6966K & \rejectcell & 6966K \\
49489, 0.25 & \acceptcell & 49K & \rejectcell & 40K & \rejectcell & 8893K & \rejectcell & 8924K & \rejectcell & 8925K & \rejectcell & 8925K \\
39387, 0.2  & \acceptcell & 47K & \rejectcell & 40K & \rejectcell & 9430K & \rejectcell & 9467K & \rejectcell & 9467K & \rejectcell & 9468K \\
77775, 0.39 & \acceptcell & 50K & \rejectcell & 42K & \rejectcell & 7474K & \rejectcell & 7318K & \rejectcell & 7318K & \rejectcell & 7318K \\
89897, 0.45 & \acceptcell & 51K & \rejectcell & 41K & \rejectcell & 6771K & \rejectcell & 6611K & \rejectcell & 6612K & \rejectcell & 6612K \\
\midrule
29285        & \acceptcell & 982K & \rejectcell & 1022K & \rejectcell & 1023K & \rejectcell & 2266K & \acceptcell & 980K & \acceptcell & 976K \\
69693        & \acceptcell & 1024K & \rejectcell & 1067K & \rejectcell & 1067K & \rejectcell & 5393K & \acceptcell & 1024K & \acceptcell & 1016K \\
100000       & \acceptcell & 1041K & \rejectcell & 1081K & \rejectcell & 1083K & \rejectcell & 7739K & \acceptcell & 1042K & \acceptcell & 1032K \\
83836        & \acceptcell & 1032K & \rejectcell & 1074K & \rejectcell & 1074K & \rejectcell & 6487K & \acceptcell & 1036K & \acceptcell & 1037K \\
53530        & \acceptcell & 1008K & \rejectcell & 1049K & \rejectcell & 1046K & \rejectcell & 4142K & \acceptcell & 1012K & \acceptcell & 1007K \\
23224        & \acceptcell & 968K & \rejectcell & 999K & \rejectcell & 1006K & \rejectcell & 1797K & \acceptcell & 963K & \acceptcell & 966K \\
\bottomrule
\end{tabular}

    \caption{\small Run of \tester{} on the original sampler (1) and its buggy variants (2–6). The Dec. column represents the outcome of \tester{}: `A' denoting an \accept{} and `R' denoting a \reject{}; the `\# Calls' column represents the number \intcond queries. The top half reports results for Binomial samplers, and the bottom half for Poisson samplers.}
    \label{tab:case-study}
\end{table*}

\begin{figure}[h]
\centering
\begin{minipage}{0.99\linewidth}
\begin{lstlisting}[
    language=Python,
    style=mystyle,
    caption={},
    literate={u_low}{{\textcolor{red}{u\_low}}}1
                {u_high}{{\textcolor{red}{u\_high}}}1
                {x_low}{{\textcolor{red}{x\_low}}}1
                {x_high}{{\textcolor{red}{x\_high}}}1
]
def Geometric(p, x_low, x_high):
    u_low = 1 - (1 - p) ** x_high
    u_high = 1 - (1 - p) ** x_low
    U = uniform(u_low, u_high) 
    return ceil(log(1 - U) / log(1 - p))
\end{lstlisting}
\end{minipage}
\caption{\small \intcond{} implementation for a standard geometric sampler. Variables introduced specifically to support interval conditioning are highlighted.}
\label{fig:geo_sampler}
\end{figure}

\subsection{Evaluation}
To evaluate the practical effectiveness of \tester{}, we implemented a prototype in Python3. The objective of our empirical evaluation was to answer the following questions: 
\textbf{RQ1:} Is \tester{} accurate and scalable for testing inverse transform samplers? \textbf{RQ2:} Is \tester{} better than the baseline algorithm?
\paragraph{Baseline and Parameters}
We implemented the algorithm \baseline proposed in \cite{bhattacharyya2024testing} and used it as a baseline for evaluating our tool. \baseline{} uses a conditioning approach called, \emph{subcube conditioning}, which is a generalization of the interval conditioning approach~\cite{CRS}. To enable compatibility with inverse transform samplers, we adapted \baseline{} with non-algorithmic modifications.
Following the setup of \cite{bhattacharyya2024testing}, we set the parameters $\eps = 0.01$ and $\eta = 0.5$ for \tester{} and \baseline{}. We set the confidence parameter $\delta = 0.1$ for both algorithms.
We conducted all our experiments on a high performance computer cluster, with each node consisting of Intel Xeon Gold 6148 CPUs. We allocated one CPU core and
a 5GB memory limit to each tester instance pair.

\paragraph{Performance Experiments} 
We conducted performance experiments to evaluate the runtime of \tester{} (in \tvident{} mode) and \baseline{}. Since \baseline{} is limited to distributions over finite domains, we restricted the performance comparison experiments to the Binomial samplers.
The benchmarks used for the performance experiments are different $n$ and $p$ parameters for the Binomial sampler, where $n$ ranges from $1{,}000$ to $600{,}000$ and $p$ ranges from $0.01$ to $0.5$. We maintained an equal split between the benchmarks that are close to the Binomial distribution and those that are far from it to ensure a balanced evaluation. 
The results of the performance experiments, shown in \Cref{fig:perf-exp}, demonstrate that \tester{} achieves an average speedup of over $1000 \times$ compared to \baseline{} across nearly all benchmarks.

\paragraph{Case Study}
We performed a case study to evaluate the effectiveness of \tester{} in detecting incorrect implementations of Binomial and Poisson samplers. We used the \infident{} mode in this setting. To simulate buggy implementations, we manually introduced errors by perturbing the constants used in their inverse sampling routines. For each distribution, we constructed six implementations: implementation 1 is correct, while implementations 2–6 contain injected bugs. For the implementation 5 and 6 of the Poisson sampler we apply benign changes by modifying the rejection sampling parameters that preserves correctness and should be accepted by our tester.
\Cref{tab:case-study} presents an abridged version of the results. \tester{} correctly identifies all samplers for almost all parameter setting that deviate from the intended distribution, while correctly accepting the implementations where the modifications preserve distributional correctness. The full experimental setup and detailed results are available in the supplementary material.

\begin{figure}[htbp!]
    \centering
\includegraphics[width=0.9\linewidth]{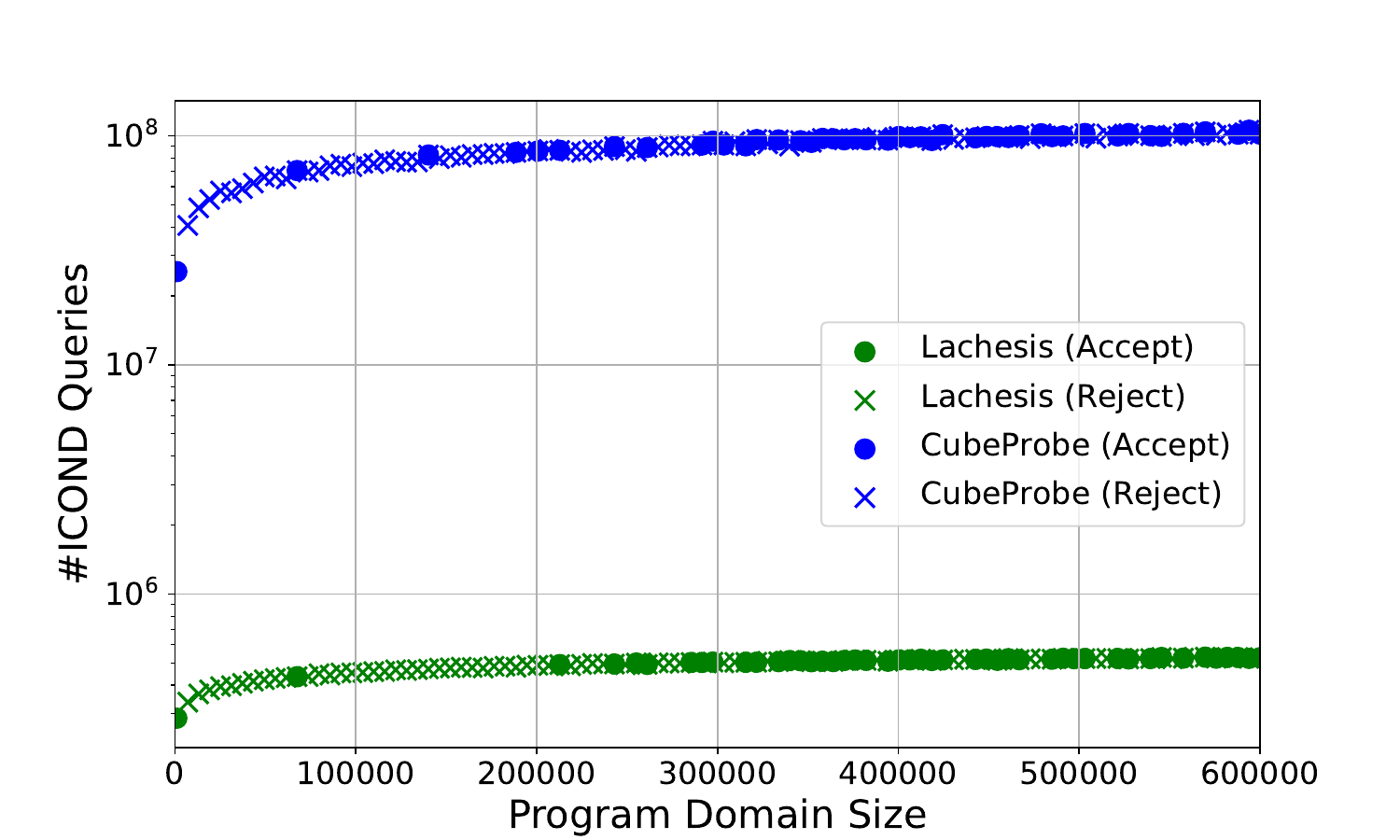}
    \caption{\small Performance comparison of \tester{} and \baseline{} on Binomial samplers. The y-axis represents the number of samples drawn from the sampler, while the x-axis shows the Domain size of the distribution.}
    \label{fig:perf-exp}
\end{figure}
\section{Conclusion}\label{sec:conclusion}
In this work, we designed the first instance-dependent tolerant testing algorithms in the interval conditioning model. Moreover, our algorithms work very well in practice, as demonstrated by our experimental results. 

\paragraph{Limitations of our work}
The query complexities of our algorithms depend on the $\tilt$ of the distributions to be tested, whereas the known lower bound of these problems is $\Omega(\log n/\log \log n)$ in the worst case~\cite{CRS}. Finding a lower bound in terms of $\tilt$ of the distributions remains an intriguing open problem.

\newpage

\section*{Acknowledgements}
The authors would like to thank the anonymous reviewers for their comments which improved the presentation of the paper.
Rishiraj Bhattacharyya acknowledges the support of UKRI
by EPSRC grant number EP/Y001680/1. Sourav Chakraborty's research is partially supported by the SERB project SQUID-1982-SC-4837. Uddalok Sarkar is supported by the Google PhD Fellowship. Sayantan Sen's research is 
supported by the NRF Investigatorship award (NRF-NRFI10-2024-0006)
and CQT Young Researcher Career Development Grant (25-YRCDG-SS). Computations were performed on the Niagara supercomputer at the SciNet HPC Consortium. SciNet is funded by Innovation, Science and Economic Development Canada; the Digital Research Alliance of Canada; the Ontario Research Fund: Research Excellence; and the University of Toronto.

\bibliography{reference}

\newpage
\appendix
\onecolumn
\section*{Supplementary Material}

\paragraph{Organization}
The supplementary material is organized as follows. We start by discussing extended preliminaries of our work, followed by the correctness proofs of our subroutines. Then we prove our main result. Finally, we conclude with details of our experimental results.

\section{Extended Preliminaries}

Let us start with the definitions of uniform distribution and Poisson random variable.

\begin{defi}[Uniform distribution]
The \emph{Uniform} distribution $\uni^{\contint{u,v}}$ is defined as 
    \[
    \uni^{\contint{u,v}}(x) = 
    \begin{cases}
        \frac{1}{|v-u|} & u \leq x \leq v \\
        0 & \text{otherwise}
    \end{cases}
    \]    
\end{defi}

\begin{defi}[Poisson random variable]
Let $\lambda >0$. A discrete random variable $X$ is said to be a \emph{Poisson} random variable with parameter $\lambda$, denoted as $\mathsf{Poi}^{\lambda}$ if the following holds:
\begin{equation}
\forall k \in \mathbb{N}, \Pr[X=k]=e^{-\lambda} \frac{\lambda^k}{k!}.
\end{equation}
\end{defi}

Now we will proceed to discuss the Tootsie Pop Algorithm (\tpa).

\subsection{Tootsie Pop Algorithm}

Before proceeding to discuss \tpa, let us define the notion of interval nests, which is crucially used in \tpa.

\paragraph{Interval Nests}
Given an element $a \in [n]$, we define the sequence of sets $\{A_a(\beta)\}$ for $0 \leq \beta \leq n$ as follows: for $x \in \contint{a-n,a+n}$, $x \in A_a(\beta)$ if and only if $\left|x - a \right| \leq \beta$. We refer to this sequence as the \emph{interval nest} with respect to the element $a$. 
The family of intervals ${A_a(\beta)}$ forms a nested structure: for any fixed $a \in [n]$, and $\beta_1 \leq \beta_2$, we have $A_a(\beta_1) \subseteq A_a(\beta_2)$. In particular, $A_a(0) = \{a\}$ and $A_a(n) = \contint{a-n, a+n}$. We define $A_a(\betacent)$ as the \emph{center} of the interval nest, such that $\betacent = \frac{1}{2}$. 
For any element $a \in [n]$, we can express $\unknown(a)$ as the ratio of the probability masses of two intervals under the augmented distribution $\cunknown{\trih}$, that is, $\unknown(a) = \tfrac{\cunknown{\trih}(A_a(\betacent))}{\cunknown{\trih}(A_a(\beta_0))}$.

\begin{restatable}{cl}{pcont}
   \label{lem:pcont}
    $\frac{\cunknown{\trih}(A_a(\betacent))}{\cunknown{\trih}(A_a(\beta_0))} = \unknown(a)$ where $\betacent = \frac{1}{2}$, $\beta_0 = n$.
\end{restatable}
\begin{proof}
Since $\betacent = \frac{1}{2}$, from the definition of interval nests, we know that  $A_a(\betacent) = \contint{a - \frac{1}{2}, a + \frac{1}{2}}$, and $A_a(\beta_0) = \contint{a - n, a + n}$. Therefore, we can say that
$$\frac{\cunknown{\trih}(A_a(\betacent))}{\cunknown{\trih}(A_a(\beta_0))} = \frac{\int_{a - \frac{1}{2}}^{a + \frac{1}{2}} \cunknown{\trih}(x)dx}{\int_{a - n}^{a + n} \cunknown{\trih}(x)dx} = \unknown(a)$$ 
This completes the proof.
\end{proof}

\begin{restatable}{lem}{nestuni}
    \label{lem:nestuni}
    Fix $a, b > 0$. Assume that we sample an element $x$ from the distribution $\cunknown{\trih}_{|A_a(b)}$ conditioned on the set $A_a(b)$. Let $b' = \inf\{\beta : x \in A_a(\beta)\}$. Then $\frac{\cunknown{\trih}(A_a(b'))}{\cunknown{\trih}(A_a(b))} \sim \uni^{\contint{0,1}}$.     
\end{restatable}
\begin{proof}
    Let us consider some $p \in [0, 1)$, and let $U := \frac{\cunknown{\trih}(A_a(b'))}{\cunknown{\trih}(A_a(b))}$. We need to show that $\Pr(U \leq p) = p$.
    Since $\cunknown{\trih}$ is a continuous function of $b$ with $\lim_{b \to 0} \cunknown{\trih}(A_a(b)) = 0$, there must exist $\beta_p \in (0,\beta]$ such that $$\frac{\cunknown{\trih}(A_a(\beta_p))}{\cunknown{\trih}(A_a(b))} = p$$
    
    Consider $0 < \eps < 1 - p$. Then there is also a $\beta_{p + \eps}$ with the same property. 
    Now consider $X \sim \cunknown{\trih}_{|A_a(b)}$. Let $b' = \inf\{\beta : X \in A_a(\beta)\}$. Because, for $\beta_1 \leq \beta_2$, we have $A_a(b_1) \subseteq A_a(b_2)$, therefore observe that, $X \in A_a(\beta_p) \implies b' \leq \beta_p \implies A_a(b') \subseteq A_a(\beta_p) \implies \cunknown{\trih}(A_a(b')) \leq \cunknown{\trih}(A_a(\beta_p)) \implies U \leq p$.  Therefore, 
    \begin{align*}
        &\Pr(X \in A_a(\beta_p)) \leq \Pr(U \leq p)\\
        \implies &~~~~p \leq \Pr(U \leq p)   
    \end{align*}
    Similarly, $X \notin A_a(\beta_{p+\eps}) \implies p + \eps \geq U$. Therefore,
    \begin{align*}
        &\Pr(X \notin A_a(\beta_{p+\eps})) \leq \Pr(p + \eps \geq U)\\
        \implies &\Pr(p + \eps \leq U) \leq \Pr(X \in A_a(\beta_{p+\eps}))\\
        \implies &\Pr(U \leq p + \eps) \leq p + \eps\\
    \end{align*}
    Thus,
    $p \leq \Pr(U \leq p) \leq \Pr(U \leq p + \eps) \leq p + \eps$. Lastly, because this argument holds for arbitrary $\eps$, we have $\Pr (U \leq p) = p$. Therefore, we can say that $U \sim \uni^{\contint{0,1}}$. 
\end{proof}

The Tootsie Pop Algorithm (\tpa)~\cite{banks2010using}, estimates the probability mass of a target interval within the domain of a continuous distribution. \tpa works by iteratively refining the interval until it reaches the target interval. 
We describe the Tootsie Pop Algorithm with a slight modification in \cref{alg:tpa}, adapted to operate using the \cintcond{} oracle. It takes as input a continuous distribution $\cunknown{\trih}$, an element $x$, number of iterations $r$, and parameters $\Thresh$ and $\delta$ which are related to the \intcond{} oracle. The algorithm returns an estimate of the probability mass of the interval $A_x(\tfrac12)$, or $\bot$ if it fails to estimate it within the given parameters. Within each iteration, the algorithm start with the initial interval $\contint{0,n}$ and refines to $\contint{x - \tfrac12, x + \tfrac12}$ by sampling from the distribution $\cunknown{\trih}$ conditioned on the current interval. It counts the number of steps ($\lambda$) taken to reach the center interval $A_x(\tfrac12)$, and returns the average number of steps taken across all iterations. If the number of steps exceeds a threshold $\Thresh$ at any iteration or if the sampling fails, it returns $\bot$. 
The following result from \citet[Lemma 2]{banks2010using} shows that the stopping time $\lambda$ follows a suitable Poisson distribution, which establishes the correctness of the algorithm \tpa{}, that is, $\lambda \sim \pois^\mu$, where $\mu = \log(\cunknown{\trih}(A_a(\beta_0))) -\log(\cunknown{\trih}(A_a(\beta_c)))$.

\begin{algorithm}[tb]
\caption{$\tpa(\cunknown{\trih}, x, r, \Thresh, \delta, \theta)$}\label{alg:tpaext}
\begin{algorithmic}[1]
    \STATE $k \gets 0$, $\beta_c \gets \frac{1}{2}$
    \FOR {$i = 1$ to $r$} \label{ln:est:for}
        \STATE $\lambda \gets 0$, $\beta \gets n$
        \WHILE { $\beta > \beta_c$}
        \STATE $u \gets \max(0,x - \beta)$, $v \gets \min(n,x + \beta)$
        \STATE $y \gets \cintcond\left(\cunknown{\trih}, u, v, \frac{\delta}{r \Thresh}, \theta\right)$
        \IF{$y = \bot$ or $\lambda\geq \Thresh$}
        \STATE {\bfseries Return} $\bot$
        \ENDIF

        \STATE $\lambda \gets \lambda + 1$, $\beta \gets |y-x|$
        \ENDWHILE
        \STATE $k \gets k + (\lambda-1)$\;
    \ENDFOR
    \STATE $\lambda \gets \frac{k}{r}$
    \STATE {\bfseries Return} $\lambda$
    \end{algorithmic}
\end{algorithm}

\begin{lem}[{\cite[Lemma 2]{banks2010using}}]
\label{lem:linf:tpapois}
    Let $\beta_c = \frac{1}{2}$. 
    Consider a sequence $\beta_0 \leq \beta_1 \leq \beta_2 \ldots$ with a stopping time $\lambda = \inf\{t \mid \beta_t \leq \beta_c\}$.
    Additionally, assume that $\frac{\cunknown{\trih}(A_a(\beta_{i+1}))}{\cunknown{\trih}(A_a(\beta_{i}))} \sim \uni^{\contint{0,1}}$.
    Then, the stopping time $\lambda$ follows a Poisson distribution, that is, $\lambda \sim \pois^\mu$ where
    $$\mu = \log(\cunknown{\trih}(A_a(\beta_0))) -\log(\cunknown{\trih}(A_a(\beta_c)))$$
\end{lem}

\subsection{Concentration Bounds}

We will use the following Chernoff bounds in our analysis. The first one is a generalization of the Chernoff bound for independent Bernoulli random variables, while the second and third ones are specific to Poisson and i.i.d\ Bernoulli random variables, respectively. See~\cite{dubhashi2009concentration} for proofs.

\begin{lem}[Chernoff Bound 1]\label{lem:cher1}
Suppose $v_1, \ldots, v_n$ are independent random variables taking values in $\{0, 1\}$. Let $V = \sum_{i=1}^nv_i$ and $\mu = \mathbb{E}[V]$ then  
\[\Pr\left(\left|V - \mu\right| \geq \eps \mu\right) \leq 2e^{-\frac{\eps^2\mu}{3}}\]
\end{lem}

\begin{lem}[Chernoff Bound 2]
    \label{lem:cher2}
    Let $x \sim \pois^\mu$ be a Poisson random variable with parameter $\mu$. Then,
    \[\Pr(x \geq \mu + a) \leq \exp\left(-\frac{a^2}{2(a + \mu)}\right), \ \text{ and } \ \Pr(x \leq \mu - a) \leq \exp\left(-\frac{a^2}{2\mu}\right)\]
\end{lem}

\begin{lem}[Chernoff Bound 3]
    \label{lem:cher3}
    Let $Y_1, Y_2, \ldots , Y_n$ be i.i.d $0-1$ random variables.
    \begin{itemize}
        \item If $\E[Y_i] \geq \theta \geq 0$ then for any $t\leq \theta$, \[\Pr\left(\frac{1}{n}\sum_{j\in[n]} Y_j\leq t\right) < \exp\left(-\frac{(\theta - t)^2n}{2\theta}\right)\]
        \item If $\E[Y_i] \leq \theta$ then for any $t\geq \theta$, \[\Pr\left(\frac{1}{n}\sum_{j\in[n]} Y_j \geq t \right) < \exp\left(-\frac{(t - \theta)^2n}{2t}\right)\]
    \end{itemize}
\end{lem}

\section{Correctness Proofs of Subroutines}
In this section, we shall prove the correctness of various subroutines. Let us start with \cintcond.

\subsection{Correctness of \cintcond Simulation}

\begin{restatable}{lem}{intcondsamp}\label{lem:intcondsamp}
If \cintcond{} returns a real number, the returned number is sampled from $\cunknown{\trih}_{|\contint{u,v}}$. 
\end{restatable}

\begin{proof}[Proof of \cref{lem:intcondsamp}]
Let $x$ be the sample drawn from $\intcond(P, [\rnd{u}, \rnd{v}])$. Since sampling of $x$ and $r$ are done independently from $\unknown_{|[\rnd{u},\rnd{v}]}$ and $\trih$, respectively, $x$ and $r$ are two independent random variables. We need to show $\Pr(u \leq x + r \leq X) = \int_{u}^X \cunknown{\trih}(x) dx$.

Consider some $X \in \contint{u,v}$. Then
    \begin{align*}
        &\Pr\left(x + r\leq X\right)\\ = &\Pr(x \leq \rnd{X} - 1) \cdot \Pr(-\frac{1}{2} \leq r \leq \frac{1}{2}) + \Pr(x = \rnd{X}) \cdot \Pr(-\frac{1}{2} \leq r \leq X - \rnd{X})\\
        =& \sum_{i=1}^{\rnd{X} - 1} \unknown(i) \cdot 1 + \unknown(\rnd{X}) \cdot \trih(r \leq X - \rnd{X})\\
        =& \sum_{i=1}^{\rnd{X} - 1} \unknown(i) + \unknown(\rnd{X}) \cdot \trih(x \leq X - \rnd{X})
    \end{align*}
    Consequently,
    $$\Pr(u \leq x + r\leq X) = \sum_{i=\rnd{u} - 1}^{\rnd{X} - 1} \unknown(i) + \unknown(\rnd{X}) \cdot \trih(x \leq X - \rnd{X}) - \unknown(\rnd{u}) \cdot \trih(x \leq u - \rnd{u})$$
    Similarly from the definition of $\cunknown{\trih}$, we can deduce that
    \begin{align*}
        &\int_{-\infty}^X \cunknown{\trih}(x) dx\\ 
        =& \int_{-\infty}^X \int_{-\infty}^{\infty} \unknown(t) \trih(x - t)  \ dt \ dx \\
        =& \sum_{i=1}^{\rnd{X} - 1} \unknown(i) \cdot \int_{i-1/2}^{i+1/2} \trih(x - i)  \ dx  + \unknown(\rnd{x}) \cdot \int_{\rnd{X}-1/2}^{X} \trih(x - \rnd{x})  \ dx\\
        =& \sum_{i=1}^{\rnd{X} - 1} \unknown(i) + \unknown(\rnd{X}) \cdot \tri(x \leq X - \rnd{X})
    \end{align*}
    Therefore, for any $X > 0$, we have $\Pr(x + r\leq X) = \int_{\infty}^X \cunknown{\trih}(x) dx$. And consequently, $\Pr(u \leq x + r\leq X) = \int_{u}^X \cunknown{\trih}(x) dx$.
\end{proof}

\begin{restatable}{lem}{intcondtime}\label{lem:intcondtime}
    Let $L = \max\left\{\tfrac{\unknown(u)}{\unknown([u+1, v])}, \tfrac{\unknown(v)}{\unknown([u, v-1])}\right\}$. Then
    \cintcond{} makes $\Oh(L)$ calls to the \intcond{} oracle in expectation. Moreover, given an upper bound $\theta \geq L$, the \cintcond{} oracle returns a sample from $\cunknown{\trih}_{|\contint{u, v}}$ with probability at least $1 - \delta$.
\end{restatable}

\begin{proof}[Proof of \cref{lem:intcondtime}]
    To analyze the expected number of calls to the \intcond{} oracle made by \cintcond{}, consider the event $\mathsf{TryAgain}$, which occurs when $x + r \notin \contint{u, v}$. The loop in \cintcond{} continues until $\neg\mathsf{TryAgain}$ occurs, so the number of iterations is a geometric random variable with success probability $\Pr(\neg\mathsf{TryAgain})$.

    Let us bound $\Pr(\mathsf{TryAgain})$. The only chance of rejecting a sample is if $x$ is at the endpoints $u$ or $v$, since for $x \in [u+1, v-1]$ the loop terminates immediately. Thus,
    \[
        \Pr(\mathsf{TryAgain}) \leq \frac{\unknown(u) + \unknown(v)}{\unknown([u, v])}
    \]
    By definition, $L = \max\left\{\frac{\unknown(u)}{\unknown([u+1, v])}, \frac{\unknown(v)}{\unknown([u, v-1])}\right\}$. Observe that
    \[
        \frac{\unknown([u, v])}{\unknown(u)} = \frac{\unknown([u+1, v]) + \unknown(u)}{\unknown(u)} = \frac{\unknown([u+1, v])}{\unknown(u)} + 1 \geq \frac{1}{L} + 1
    \]
    and similarly for $\unknown(v)$. Therefore,
    \[
        \Pr(\mathsf{TryAgain}) \leq \frac{\unknown(u) + \unknown(v)}{\unknown([u, v])} \leq \frac{2L}{2L+1}
    \]
    This means the expected number of iterations is at most $2L+1 = \Oh(L)$.

    For the second part, suppose we run the loop for $\stp$ iterations. The probability that all $\stp$ samples are rejected is at most $\left(\Pr(\mathsf{TryAgain})\right)^{\stp}$. Given an upper bound $\theta \geq L$, if we set $\stp \geq (2\theta+1)\log\frac{1}{\delta}$, then
    \[
        \left(\frac{2\theta}{2\theta+1}\right)^{\stp} \leq \delta
    \]
    Thus, with probability at least $1-\delta$, the algorithm returns a sample from $\cunknown{\trih}_{|\contint{u, v}}$ within $\stp$ iterations. This completes the proof.

\end{proof}

\section{Correctness of \Est}

\est*

We will prove the above lemma in the following two claims.

\begin{cl}
 Given \intcond access to $\unknown$, an element $x \in [n]$, and parameters $\zeta, \delta \in (0,1)$, the algorithm \Est returns a value $\widehat{\unknown}_x$. If $\unknown(x) \geq \theta$, then with probability at least $1 - \delta$, $\widehat{\unknown}_x \in (1 \pm \zeta)\cdot \unknown(x)$ holds.     
\end{cl}

\begin{proof}
    A success in the 1st and 2nd phase of \Est algorithm is defined using the following events:
    \begin{itemize}
        \item $E_1$: The first phase of \tpa returns a value $\lambda$ such that $\lambda + \sqrt{\lambda} + 2 \geq \log \frac{1}{\unknown(x)}$.
        \item $E_2$: The second phase of \tpa returns a value $\lambda$ such that $|\lambda - \log \frac{1}{\unknown(x)}| \leq \log(1+\zeta) $. 
    \end{itemize} 
    We begin with the first phase of \Est, where \tpa is run with $r_1 = 2 \log \frac{8}{\delta}$, and analyze the event $E_1$.  
    The event $E_1$ always holds if $L \leq 2$, where $L = \log \tfrac{1}{\unknown(x)}$.  
    When $L > 2$, the event $E_1$ holds if the returned estimate $\lambda$ satisfies
    \[
    \lambda \geq L - \tfrac{3}{2} - \sqrt{L - \tfrac{7}{4}}.
    \]

    By \cref{lem:linf:tpapois}, we have $\lambda \sim \pois^L$. Consequently, $r_1 \lambda \sim \pois^{r_1 L}$, and thus we can apply a standard Poisson tail bound (Lemma~\ref{lem:cher2}) to obtain:
    \begin{align*}
    \Pr(\overline{E_1})
    &= \Pr\left(\lambda < L - \tfrac{3}{2} - \sqrt{L - \tfrac{7}{4}}\right) \\
    &= \Pr\left(\pois^{r_1 L} < r_1 L - r_1\left(\tfrac{3}{2} + \sqrt{L - \tfrac{7}{4}}\right)\right) \\
    &\leq \exp\left(-\frac{1}{2} \cdot \frac{\left(r_1 \left(\tfrac{3}{2} + \sqrt{L - \tfrac{7}{4}}\right)\right)^2}{r_1 L}\right) \\
    &= \exp\left(-\frac{r_1}{2} \cdot \frac{\left(\tfrac{3}{2} + \sqrt{L - \tfrac{7}{4}}\right)^2}{L}\right) \\
    &\leq \exp\left(-\tfrac{1}{2} r_1\right) \qquad \qquad \qquad \qquad \qquad \qquad \text{(since } \left(\tfrac{3}{2} + \sqrt{L - \tfrac{7}{4}}\right)^2 \geq L \text{ for } L > 2\text{)} \\
    &= \exp\left(-\log \tfrac{8}{\delta}\right) = \tfrac{\delta}{8}.
    \end{align*}

    Now, suppose the event $E_1$ occurs, and consider the second phase. Since $E_1$ holds, we have
    \[
    r_2 \geq \frac{2 \ln(16/\delta)}{(L + \ln(1 + \zeta))^2}.
    \]
    We analyze the probability that the estimate $\lambda$ deviates from $L$ by more than $\ln(1 + \zeta)$.

    Using \cref{lem:cher2}, we obtain an upper tail bound:
    \begin{align*}
    \Pr\left(\lambda \geq L + \ln(1 + \zeta) \mid E_1\right)
    &= \Pr\left(r_2 \lambda \geq r_2 L + r_2 \ln(1 + \zeta) \mid E_1\right) \\
    &\leq \exp\left(-\frac{1}{2} \cdot \frac{(r_2 \ln(1 + \zeta))^2}{r_2 L + r_2 \ln(1 + \zeta)}\right) \\
    &= \exp\left(-\frac{r_2 \ln^2(1 + \zeta)}{2(L + \ln(1 + \zeta))}\right) \\
    &\leq \exp\left(-\ln \tfrac{4}{\delta}\right) = \tfrac{\delta}{16}.
    \end{align*}

    Similarly, for the lower tail:
    \begin{align*}
    \Pr\left(\lambda \leq L - \ln(1 + \zeta) \mid E_1\right)
    &= \Pr\left(r_2 \lambda \leq r_2 L - r_2 \ln(1 + \zeta) \mid E_1\right) \\
    &\leq \exp\left(-\frac{1}{2} \cdot \frac{(r_2 \ln(1 + \zeta))^2}{r_2 L}\right) \\
    &= \exp\left(-\frac{r_2 \ln^2(1 + \zeta)}{2L}\right) \\
    &\leq \tfrac{\delta}{16}.
    \end{align*}

    By the union bound over failure of $E_1$ and the two deviations in $E_2$, the total failure probability is at most
    \[
    \Pr(\overline{E_1}) + \Pr(\overline{E_2} \mid E_1) \leq \tfrac{\delta}{8} + \tfrac{\delta}{16} + \tfrac{\delta}{16} = \tfrac{\delta}{4}.
    \]

    Finally, if $r_2 \lambda$ is within additive error $r_2 \ln(1 + \zeta)$ of $L$, then $e^{-\lambda}$ is within a multiplicative factor of $1 + \zeta$ of $\unknown(x)$, concluding the proof.
\end{proof}

\begin{cl}\label{cl:estquery_app}
The expected number of \intcond queries performed by \Est is $\tOh\left(\frac{1}{(\eta - \eps)^2} \E_{x\sim \unknown}\left[\min(\tilt_\unknown(x), \theta)\cdot \log^2 \tfrac{1}{\unknown(x)}\right]\right)$
\end{cl}

We will need the following claim to prove the query complexity of \Est.

\begin{cl}\label{cl:pois}
Suppose the random variable $X$ follows a Poisson like distribution $\pois'^{(\mu, M)}$, such that $\pois'^{(\mu,M)}(x) \propto \pois^\mu(x)$ for $x \leq M$, and $\pois'^{(\mu,M)}(x) = 0$ for $x > M$. Then $\E[X^2] \leq \min(\mu^2 + \mu , M^2)$.
\end{cl}
    
\begin{proof} 
Let $\sum_{x = 0}^{M} \frac{\mu^xe^{-\mu}}{x!} = \Delta$. For the expectation of $X$, we have
\begin{align*}
    \E[X] = \sum_{x = 0}^{M} x \cdot \frac{\mu^xe^{-\mu}}{x!} \cdot \frac{1}{\Delta}
    = \frac{e^{-\mu}}{\Delta} \sum_{x = 1}^{M} x \cdot \frac{\mu^x}{x!}
    = \frac{\mu e^{-\mu}}{\Delta} \sum_{x = 0}^{M-1} \frac{\mu^{x}}{x!}
\end{align*}
Since, $\sum_{x = 0}^{M-1} \frac{\mu^x}{x!} \leq \sum_{x = 0}^{M} \frac{\mu^x}{x!} = \Delta e^{\mu}$, it follows that:
\begin{align*}
    \E[X] \leq \frac{\mu e^{-\mu}}{\Delta} (\Delta e^{\mu}) \leq \mu
\end{align*}
Similarly, for $\E[X(X-1)]$ we have:
\begin{align*}
    \E[X(X-1)] &= \sum_{x = 0}^{M} x(x-1) \cdot \frac{\mu^xe^{-\mu}}{x!} \cdot \frac{1}{\Delta} = \frac{e^{-\mu}}{\Delta} \sum_{x = 2}^{M} \frac{\mu^x}{(x-2)!}
    = \frac{\mu^2 e^{-\mu}}{\Delta} \sum_{x = 2}^{M} \frac{\mu^{x-2}}{(x-2)!}\\
\end{align*}
Similarly, since $\sum_{x = 0}^{M-2} \frac{\mu^x}{x!} \leq \sum_{x = 0}^{M} \frac{\mu^x}{x!} = \Delta e^\mu$, we get:
\begin{align*}
    \E[X(X-1)] \leq \frac{\mu^2 e^{-\mu}}{\Delta} (\Delta e^{\mu}) = \mu^2
\end{align*}

Thus, $\E[X^2] = \E[X(X-1)] + \E[X] \leq \mu^2 + \mu$. Again, since $x \leq M$, it follows that, $\E[X(X-1)] \leq M^2-M$ and $\E[X] \leq M$, so $\E[X^2] \leq M^2$. Therefore, $\E[X^2] \leq \min(\mu^2 + \mu , M^2)$. 
\end{proof}

\begin{lem}
\label{lem:tiltest}
Given an input $x$ to \tpa, the expected number of \intcond queries performed by \cintcond is at most $\Oh(\min(\tilt_\unknown(x), \theta))$.
\end{lem}
\begin{proof}
    By observing that for any $u, v$ such that $x \in [u, v]$, we have from \cref{lem:intcondtime} that the expected number of \intcond queries performed by \cintcond is at most:
    \begin{align*}
        L & \leq \max\left\{\tfrac{\unknown(u)}{\unknown[u+1, v]}, \tfrac{\unknown(v)}{\unknown[u, v-1]}\right\}\\
        & \leq \max\left\{\tfrac{\unknown(u)}{\unknown[u+1, x]}, \tfrac{\unknown(v)}{\unknown[x, v-1]}\right\}\\
        & \leq \max_{y\in [n]}\left\{\tfrac{\unknown(y)}{\unknown[y+1, x]}, \tfrac{\unknown(y)}{\unknown[x, y-1]}\right\} = \tilt_\unknown(x)
    \end{align*}
    The second inequality follows from the fact that $\unknown[u+1, v] \geq \unknown[u+1, x]$ and $\unknown[u, v-1] \geq \unknown[x, v-1]$ and the last inequality follows from taking maximum over all $y \in [n]$.
\end{proof}

\begin{proof}[Proof of \cref{cl:estquery_app}]
    Note that \Est calls \tpa twice with parameters $r_1,r_2$, where $r_1=2\log \frac{8}{\delta_\Est}$ and $r_2=2(\lambda + \sqrt{\lambda} +2 + \log(1+\zeta)) \frac{1}{\log^2 (1+\zeta)} \log \frac{16}{\delta_\Est}$. Observe that $\lambda$ follows a Poisson like distribution with mean $\log \frac{1}{P(x)}$ and clipped at $\Thresh$. Therefore, from \cref{cl:pois}, for a fixed $x$, $\E[\lambda^2 \mid x] = \Oh\left(\min\left(\log^2 \frac{1}{P(x)}, \Thresh^2\right)\right)$. From the description of the algorithm \tpa, we know that \tpa calls \cintcond $r_1 \lambda$ and $r_2 \lambda$ times in expectation (whenever it is called with parameters $r_1$ and $r_2$ respectively). Finally, from \Cref{lem:tiltest}, we know that each call to \cintcond is simulated by at most $\Oh(\min(\tilt_\unknown(x), \theta))$. Thus, the expected query complexity of \Est is given by:
\begin{eqnarray*}
&&\tOh\left(\E_{x \sim \unknown}\left[ \E\left[(r_1 +r_2) \cdot \lambda \cdot \min(\tilt_\unknown(x), \theta) \mid x\right] \right]\right)\\
&=& \tOh\left(\E_{x \sim \unknown}\left[ \E\left[ \left(2\log \frac{8}{\delta_\Est} + 2(\lambda + \sqrt{\lambda} +2 + \log(1+\zeta)) \frac{1}{\log^2 (1+\zeta)} \log \frac{16}{\delta_\Est}\right) \cdot \lambda \cdot \min(\tilt_\unknown(x), \theta) \mid x\right]\right]\right)\\
&=& \tOh\left(\frac{1}{\log^2(1 + \frac{2}{\eta - \eps + 2})} \E_{x\sim \unknown}\left[\min(\tilt_\unknown(x), \theta) \cdot \E[\lambda^2 \mid x]\right]\right)\\
&=& \tOh\left(\frac{1}{\log^2(1 + \frac{2}{\eta - \eps + 2})} \E_{x\sim \unknown}\left[\min(\tilt_\unknown(x), \theta)\cdot \min\left(\log^2 \frac{1}{\unknown(x)}, \Thresh^2\right)\right]\right)\\
&=& \tOh\left(\frac{1}{(\eta - \eps)^2} \E_{x\sim \unknown}\left[\min(\tilt_\unknown(x), \theta)\cdot \min\left(\log^2 \frac{1}{\unknown(x)}, \Thresh^2\right)\right]\right)
\end{eqnarray*}

The last equality follows from the fact that $\log^2(1 + \frac{2}{\eta - \eps + 2}) = \Theta((\eta - \eps)^2)$. 
This completes the proof.
\end{proof}
\section{Proof of our main result: \Cref{thm:tvident}}

In this section, we will prove our main technical result.

\maintvident*

We will separate the proof of the above theorem into two parts. We will first prove the correctness of \tvident and then proceed to prove the correctness of \infident.

\subsection{Analysis of \tvident}

\begin{theo}\label{theo:tvindentproof_app}
Let $\unknown$ be an unknown distribution and $\known$ a known distribution over $\mathbb{Z}$. 
    Given access to $\intcond(\unknown)$, accuracy parameters $\eps, \eta \in (0,1)$ with $\eta > \eps$, and confidence parameter $\delta \in (0,1)$, the algorithms $\tvident$ $\tvident$ can distinguish between the cases $\dtv(\unknown, \known) \leq \eps$ and $\dtv(\unknown,\known) \geq \eta$ with probability $\geq 1- \delta$. The expected number of $\intcond$ queries is
    \[\tOh\left(\frac{1}{(\eta - \eps)^4} \cdot \E_{x \sim \avgknown} \left[\tilt_{\avgknown}(x) \log^2 \frac{1}{\avgknown(x)} \right] \right).\]
 where $\avgknown$ is the distribution defined by $\avgknown(x)=(P(x)+Q(x))/2$.
    
\end{theo}

To prove \cref{theo:tvindentproof_app}, we will divide the proof into three parts: (1) completeness and (2) soundness, and (3) the expected number of calls to \intcond. Finally, we will combine the results of these three parts to prove \cref{thm:tvident}. But before we proceed to the proof, we will prove the following lemma that shows that $\avgknown$ is a good approximation of $\unknown$. Fix $\eps' = \frac{\eps}{2}$ and $\eta' = \frac{\eta}{2}$ throughout this section. 

\begin{lem}\label{lem:avgdist_app}
    $\dtv(\avgknown, \known) = \frac{1}{2}\dtv(\unknown, \known)$.
\end{lem}
\begin{proof}
    $\avgknown(x) = \frac{P(x) + Q(x)}{2}$. Therefore, 
    \begin{align*}
        \dtv(\avgknown, \known) &= \frac{1}{2}\sum_{x \in [n]} |\avgknown(x) - \known(x)|\\
        &= \frac{1}{2}\sum_{x \in [n]} \left|\frac{\unknown(x) + \known(x)}{2} - \known(x)\right|\\
        &= \frac{1}{2}\sum_{x \in [n]} \left|\frac{\unknown(x) - \known(x)}{2}\right|\\
        &= \frac{1}{2}\dtv(\unknown, \known)
    \end{align*}
    This completes the proof of the lemma.
\end{proof}

\subsubsection*{Completeness of \tvident}
We will prove the following lemma.

\begin{lem}
\label{lem:tvtvcomplete}
Let $\dtv(\unknown, \known)\leq \eps$. Then with probability at least $1-\delta$, our algorithm \tvident outputs \accept. 
\end{lem}

In order to prove this lemma, let us consider the following partition of the domain $[n]$. We will partition the domain of $\avgknown$ into the following two sets: 
\begin{itemize}
    \item[(1)] $\Bad := \{x \ : \ |\avgknown(x) - \known(x)| > 3\eps' \avgknown(x)\}$,

    \item[(2)] $\Good := \{x \ : \ |\avgknown(x) - \known(x)| \leq 3\eps' \avgknown(x)\}$.
\end{itemize}

Following the above definitions of \Good and \Bad sets, we have the following claim, which bounds the total probability mass of all bad elements when $\unknown$ and $\known$ are $\eps$-close.

\begin{cl}\label{cl:badset}
Let us assume that $\dtv(\unknown, \known) \leq \eps$. Then $\sum_{x \in \mathsf{Bad}}\avgknown(x) \leq \frac{2}{3}$.    
\end{cl}

\begin{proof}
We will prove this by contradiction. Let us assume that $\sum_{x \in \mathsf{Bad}}\avgknown(x) > \frac{2}{3}$. Then we have the following:
\begin{align*}
        \dtv(\avgknown, \known) &= \frac{1}{2}\sum_{x \in [n]} |\avgknown(x) - \known(x)|\\
        &\geq \frac{1}{2}\sum_{x \in \mathsf{
        Bad}} |\avgknown(x) - \known(x)|\\  
        &> \frac{3\eps'}{2}  \sum_{x \in \mathsf{
        Bad}} \avgknown(x) && [\text{From definition of \Bad set}]\\
        &> \frac{3\eps'}{2} \cdot \frac{2}{3} && [\text{From our assumption}]\\
        &= \eps'
    \end{align*}
Note that when $\dtv(\unknown, \known) \leq \eps$, from \Cref{lem:avgdist_app}, this implies that $\dtv(\avgknown, \known) \leq \eps/2 = \eps'$. However, the above contradicts our assumption that $\dtv(\unknown, \known) \leq \eps$. This completes the proof of the claim.     
\end{proof}

Let us now define the event corresponding to the event that not enough samples from the set \Good are obtained in the multi-set $\samples$.
$$\Tiny:= \mbox{The number of samples from the set \Good is less than } t.$$

We will prove the following claim that bounds the probability of the event \Tiny.
\begin{cl}
    Suppose we have obtained a multiset of $t'=3kt$ samples from $\avgknown$. Then $\Pr(\Tiny) \leq \frac{\delta}{4}$.
\end{cl}

\begin{proof}
Consider $k=1 + \frac{1}{t}\log \frac{4}{\delta} + \sqrt{\frac{1}{t}\log^2 \frac{4}{\delta} + \frac{2}{t}\log \frac{4}{\delta}}$ as chosen in the algorithm \tvident. Following \Cref{cl:badset}, we know that $\sum_{x \in \mathsf{Good}}\avgknown(x) > \frac{1}{3}$. Now let us consider the following random variable:
$$X= \mbox{number of elements drawn that belong to } \Good$$
Thus we can say that $\E[X] \geq \frac{1}{3} \cdot 3kt = kt \geq t$. Using the Chernoff bound, we can say that 
\[\Pr(X \leq t)  = \Pr(X \leq \frac{1}{k} \cdot kt) \leq \Pr(X \leq \frac{1}{k} \cdot \E[X]) \leq e^{-\frac{(kt - t)^2}{2kt}} \leq \frac{\delta}{4}\] 
This concludes the proof.    
\end{proof}

Now we will bound the probability of the event $\Fail_\tpa$.

\begin{cl}\label{cl:tpafilprob}
Consider an element $x\in \Good$. Then $\Pr(\Fail_\tpa)\leq \delta_\tpa$.    
\end{cl}

\begin{proof}
Following the definition of \Good set, we know that if an element $x \in \Good$ then the following holds:
$$(1-3\eps)\avgknown(x) \leq \known(x) \leq (1+3\eps)\avgknown(x)$$
Therefore, similar to \cref{lem:tpafail}, we can show that $\Pr(\Fail_\tpa)\leq \delta_\tpa$.
\end{proof}

Now we show that the size of set $\cX$ following the execution of \Est is at least $t$ with high probability. 
\begin{cl}
With probability at least $1 - \delta_\tpa$, $|\samples| \geq t$ holds.   
\end{cl}

\begin{proof}
From the description of \Est (\Cref{alg:est}) and \tpa (\Cref{alg:tpa}), we know that \Est returns $\bot$ when \tpa returns $\bot$. Since $\Fail_\tpa$ denotes the event when \tpa returns $\bot$, we can say that the probability that \Est returns $\bot$ is bounded by the probability of the event $\Fail_\tpa$, which is bounded by $\delta_\tpa$ following \Cref{cl:tpafilprob}. From the description of \tvident, we can thus say that with probability at least $1 - \delta_\tpa$, $|\samples| \geq t$ holds.   
\end{proof}

Now we recall a result from \cite{banks2010using}.

\begin{cl}[{\cite[Theorem 3]{banks2010using}}]\label{lem:est_app}
For an arbitrary element $x \in [n]$, $\Pr(\Err^x_\Est \ | \ \neg \Fail^x_\Est) \leq \frac{\delta_\Est}{2}$ holds.
\end{cl}

The following claim bounds the probability of the event $(\Err_{\Est} | \neg \Fail_{\Est})$.

\begin{cl}\label{cl:tpa-huber}
Consider any element $x\in [n]$. Then $\Pr(\Err_{\Est} | \neg \Fail_{\Est}) \leq \frac{\delta_\Est}{2}$.

\end{cl}

\begin{proof}
The proof follows in a similar line from \cref{lem:est_app} and is skipped.
\end{proof}

\begin{cl}
Consider an element $x\in \Good$. Then $\Pr(\Err_{\Est}) \leq \delta_\Est$.

\end{cl}

\begin{proof}
The proof follows from \cref{lem:esterr} and is skipped.
\end{proof}

\begin{cl}\label{lem:tvtvbound}
Consider the case when $\dtv(\unknown, \known) \leq \eps$. Conditioned on the fact that the event \Tiny has not occurred,  with probability at least $1-\frac{\delta}{2}$, \tvident computes $\dest$ such that $|\dest - \dtv(\unknown, \known)| \leq \frac{\eta - \eps}{2}$.

\end{cl}

\begin{proof}
Since the event $\Tiny$ has not occurred, therefore, we know that $|\samples| > t$. Hence, similar to \cref{lem:inftvbound}, we can say that:
\begin{eqnarray*}
\Pr\left(|\dest - \dtv(\avgknown, \known)| \mid \neg \Tiny \leq \frac{\eta' - \eps'}{2}\right)) \leq \frac{\delta}{2}    
\end{eqnarray*}

\end{proof}

Now we are ready to prove the completeness of our algorithm \tvident.

\begin{proof}[Proof of \cref{lem:tvtvcomplete}]
    
Combining the above claims, we conclude that if $\dtv(\unknown, \known) \leq \eps$, that is, $\dtv(\avgknown, \known) \leq \eps'$ then 
\begin{align*}
    \Pr\left(\mbox{\tvident outputs } \reject\right) & \leq \Pr\left(\dest > \frac{\eta + \eps}{4} | \neg \Tiny\right) + \Pr(\Tiny)\\
    &\leq \frac{\delta}{2} + \frac{\delta}{4} \\
    &< \delta
\end{align*}

Therefore, with probability at least $1-\delta$, our algorithm \tvident returns \accept. This completes the proof.
\end{proof}

\subsubsection*{Soundness of \tvident}
Let us now move towards proving the soundness of our algorithm. In particular, we will prove the following lemma.

\begin{lem}
\label{lem:tvtvsound}
If $\dtv(\unknown, \known) \geq \eta$, then with probability at least $1-\delta$, our algorithm \tvident outputs \reject.    
\end{lem}

\begin{proof}
Since $\dtv(P,Q)\geq \eta$, that is, $\dtv(\unknown, \known) \geq\eta'$, we can say that
\begin{align*}
    \Pr\left(\dest < \frac{\eta+\eps}{4} | \neg \Tiny\right) &\leq \Pr\left(\dest < \dtv(\avgknown, \known) - \frac{\eta'-\eps'}{2} | \neg \Tiny\right)\\
    &\leq \frac{\delta}{2}
\end{align*}
    
Therefore, we can say that 
\begin{align*}
    \Pr(\mbox{\tvident outputs } \reject) &= \Pr\left(\dest > \frac{\eta+\eps}{4} | \neg \Tiny\right) \cdot \Pr(\neg \Tiny) + \Pr(\Tiny)\\
    &\geq (1-\delta)(1-\Pr(\Tiny)) + \Pr(\Tiny)\\
    &\geq 1 - \delta
\end{align*}
This completes the soundness proof.
\end{proof}

\subsubsection*{Query Complexity of \tvident}
The query complexity of \tvident is similar to that of \infident, and is given by the following lemma.

\begin{lem}\label{lem:tvtvquery}
The total query complexity of \tvident is $\tOh\left(\frac{1}{(\eta - \eps)^4} \cdot \E_{x \sim \avgknown} \left[\tilt_{\avgknown}(x) \log^2 \frac{1}{\avgknown(x)} \right] \right)$.
\end{lem}

\begin{proof}
    From \cref{cl:estquery_app}, we know that the expected query complexity of \Est when the input distribution is $\avgknown$ is
    \[\tOh\left(\frac{1}{(\eta - \eps)^2} \cdot \E_{x \sim \avgknown} \left[\tilt_{\avgknown}(x) \log^2 \frac{1}{\avgknown(x)} \right] \right).\]
    Since \tvident calls \Est $t$ times where $t=\frac{8}{(\eta - \eps)^2} \log \frac{4}{\delta}$, the expected query complexity of \tvident becomes
    \[\tOh\left(\frac{1}{(\eta - \eps)^4} \cdot \E_{x \sim \avgknown} \left[\tilt_{\avgknown}(x) \log^2 \frac{1}{\avgknown(x)} \right] \right).\]
\end{proof}

\begin{proof}[Proof of \cref{thm:tvident}]
    The proof follows from \cref{lem:tvtvcomplete}, \cref{lem:tvtvsound}, and \cref{lem:tvtvquery}. 
\end{proof}

\subsection{Analysis of \infident}

Now we prove the correctness of our early reject tester \infident, as stated below.

For the ease of understanding, we will first prove the correctness of our early reject tester \infident, and defer the proof of the correctness of \tvident to the next section. 

\begin{theo}\label{theo:infidentproof_app}
 Let $\unknown$ be an unknown distribution and $\known$ a known distribution over $\mathbb{Z}$. 
    Given access to $\intcond(\unknown)$, accuracy parameters $\eps, \eta \in (0,1)$ with $\eta > \eps$, and confidence parameter $\delta \in (0,1)$, the algorithm $\infident$  can distinguish between the cases $\linf(\unknown, \known)\leq 2\eps$ and $\dtv(\unknown,\known) \geq \eta$ with probability $\geq 1- \delta$. The expected number of oracle queries made by the $\infident$ is at most
\[\tOh\left(\frac{1}{(\eta - \eps)^4} \E_{x\sim \unknown}\left[\tilt_\known(x) \cdot \min\left(\log^2 \frac{1}{\unknown(x)}, \log^2 \frac{1}{\known(x)}\right)\right]\right).\]   
\end{theo}

Similar to \cref{theo:tvindentproof_app}, we will prove the above theorem in three parts: (1) Completeness of \infident, (2) Soundness of \infident, and (3) Analysis of the expected number of \intcond{} queries made by \infident.

\subsubsection*{Completeness of \infident}
Let us start with proving the completeness of our algorithm. For clarity of representation, we set $\eps_1=2 \eps$ throughout this section. Particularly, we will prove the following lemma to establish the completeness.

\begin{lem}
    \label{lem:completeness}
Let $\linf(\unknown, \known)\leq \eps_1$. Then our algorithm \infident outputs \accept with probability at least $1-\delta$.
\end{lem}

We will prove this lemma using a couple of claims. Before starting the proof, we define the following sets of events:
\begin{itemize}
    \item[(1)] $\Fail_\tpa^x$ := $\tpa$ returns $\bot$ for some $x \in \samples$.

    \item[(2)] $\Fail_\Est^x$ := $\Est$ returns $\bot$ for some $x \in \samples$.

    \item[(3)] $\Err_\Est^x$ := Either $\pest{x} \notin [(1-\zeta)\unknown(x), (1+\zeta)\unknown(x)]$ or $\Fail_\Est^x$ holds.

\end{itemize}

Let us start by bounding the probability of the event $\Fail_\tpa^x$. But even before that we show that for any $x$ such that $\unknown(x) > 0$ and $\known(x) > 0$, $\tilt_\unknown(x)$ is $\Oh(\tilt_\known(x))$.

\begin{cl}
    \label{lem:smooth}
    If $\ell_\infty(\unknown, \known) \leq \eps_1$, then for all $x \in [n]$ with $\unknown(x), \known(x) > 0$, we have $\tilt_\unknown(x)=\Oh(\tilt_\known(x))$.
\end{cl}

\begin{proof}
Since $\ell_\infty(\unknown, \known) \leq \eps_1$, we have $(1-\eps_1)\known(x) \leq \unknown(x) \leq (1+\eps_1)\known(x)$ for all $x \in [n]$. Therefore for all $y$ we have
\[\frac{\unknown(y)}{\unknown[y+1, x]} \leq (1+\eps_1)(1+\eps_1)^{-1}\cdot\frac{\known(y)}{\known[y+1, x]} = \left(\frac{1+\eps_1}{1-\eps_1}\right)\frac{\known(y)}{\known[y+1, x]}\] 
and similarly $\frac{\unknown(y)}{\unknown[x, y-1]} \leq \left(\frac{1+\eps_1}{1-\eps_1}\right)\frac{\known(y)}{\known[x, y-1]}$.
Let $y^*$ be the point which maximizes $\tilt_\unknown(x)$, i.e., $\tilt_\unknown(x) = \max\left(\frac{\unknown(y^*)}{\unknown[y^*+1, x]}, \frac{\unknown(y^*)}{\unknown[x, y^*-1]}\right)$. Using the above inequalities, we have
\[\tilt_\unknown(x) \leq \left(\frac{1+\eps_1}{1-\eps_1}\right)\max\left(\frac{\known(y^*)}{\known[y^*+1, x]}, \frac{\known(y^*)}{\known[x, y^*-1]}\right) \leq \left(\frac{1+\eps_1}{1-\eps_1}\right)\tilt_\known(x).\]
Therefore, $\tilt_\unknown(x) = \Oh(\tilt_\known(x))$.
\end{proof}

\begin{cl}\label{lem:tpafail}
Consider an arbitrary element $x \in [n]$ such that $(1-\eps_1)\unknown(x) \leq \known(x) \leq (1+\eps_1)\unknown(x)$. Then $\Pr(\Fail^x_\tpa)\leq \delta_\tpa$ holds.
\end{cl}

\begin{proof}
We begin by observing that \tpa can return $\bot$ in two cases. 

\textbf{Case 1} (When $\cintcond$ returns $\bot$): Since from \cref{{lem:smooth}}, $\tilt_\unknown(x) = \left(\frac{1+\eps_1}{1-\eps_1}\right)\tilt_\known(x)$ therefore, from \cref{lem:intcondtime}, the probability that $\cintcond$ returns $\bot$ is at most $\left(\frac{2\tilt_\unknown(x)}{2\tilt_\unknown(x)+1}\right)^T \leq \frac{\delta'_\tpa}{2\Thresh}$. Therefore, by union bound, the probability that $\cintcond$ returns $\bot$ in any of the maximum $\Thresh$ iterations is at most $\frac{\delta'_\tpa}{2}$.

\textbf{Case 2} (When $\lambda \geq \Thresh$): 
Since $(1-\eps_1)\unknown(x) \leq \known(x) \leq (1+\eps_1)\unknown(x)$, we have $\log\frac{1}{\unknown(x)} \leq \log\frac{1+\eps_1}{\known(x)}$. Let $\beta_{i-1}$ and $\beta_{i}$ be two consecutive $\beta$'s in the execution of the while loop. Then using \cref{lem:nestuni} $\frac{\cunknown{\trih}(A_x(\beta_i))}{\cunknown{\trih}(A_x(\beta_{i-1}))} \sim \uni^{\contint{0,1}}$.
Therefore from \cref{lem:linf:tpapois}, after the while loop finishes $\lambda \sim \pois^\mu$, where $\mu=(\log(\cunknown{\trih}(A_x(\beta_0))) -\log(\cunknown{\trih}(A_x(\beta_c))))$. Therefore, $\E[\lambda] = \log \frac{\cunknown{\trih}(A_x(\beta_0))}{\cunknown{\trih}(A_x(\beta_c))}$. Consequently, using \cref{lem:pcont},
$\E[\lambda] = \log\frac{1}{\unknown(x)}$. Since, $\known(x) \leq (1+\eps_1)\unknown(x)$, $\E[\lambda] \leq \log\frac{1+\eps_1}{\known(x)} = B$. 
\begin{align*}
    \Pr(\Fail^x_\tpa) &= \Pr(k \geq \Thresh + B)\\
        &\leq \Pr\left(k \geq \Thresh + \log \frac{1}{\unknown(x)}\right) && \left[\text{Because $B \geq \log\frac{1}{\unknown(x)}$}\right]\\
        &\leq \exp\left(-\frac{\Thresh^2}{2(\Thresh+ \log\frac{1}{\unknown(x)})}\right) &&\left[\text{From } \text{\cref{lem:cher2}}\right]\\
        &\leq \exp\left(-\frac{\Thresh^2}{2(\Thresh+ B)}\right) &&\left[\text{Because $B \geq \log\frac{1}{\unknown(x)}$}\right]\\
        &= \delta'_\tpa
\end{align*}
The last equality holds as $\Thresh = \log\frac{2}{\delta'_\tpa} + \sqrt{\log^2 \frac{2}{\delta'_\tpa} + 2B\log \frac{2}{\delta'_\tpa}}$. By the union bound, the probability that \tpa{} returns $\bot$ in any of the $r$ iterations is at most $r\delta'_\tpa = \delta_\tpa$. This completes the proof.
\end{proof}

Now we will bound the probability of the event $\Fail_\Est^x$.

\begin{cl}\label{lem:estfail}
Consider an arbitrary element $x \in [n]$ such that  $(1-\eps_1)\unknown(x) \leq \known(x) \leq (1+\eps_1)\unknown(x)$. Then $\Pr(\Fail^x_\Est)\leq \frac{\delta_\Est}{2}$ holds.
\end{cl}

\begin{proof}
From the description of the algorithm \Est, we know that the algorithm \tpa is called twice during the execution of \Est. Therefore by the Union bound and \cref{lem:tpafail}, we can say that the probability that \Est returns $\bot$ for $x$ is at most $2\delta_\tpa = 2\cdot\frac{\delta_\Est}{4} = \frac{\delta_\Est}{2}$. The equality follows since $\delta_\tpa=\frac{\delta_\Est}{4}$.
\end{proof}

Now we will bound the probability of the event $\Err_\Est^x$ using \Cref{lem:estfail} and \Cref{lem:est_app}.

\begin{cl}\label{lem:esterr}
Consider an arbitrary element $x \in [n]$ such that $(1-\eps_1)\unknown(x) \leq \known(x) \leq (1+\eps_1)\unknown(x)$. Then $\Pr(\Err^x_\Est) \leq \delta_\Est$. 
\end{cl}

\begin{proof}
From \cref{lem:estfail} and \cref{lem:est_app}, we have $\Pr(\Fail^x_\Est) \leq \frac{\delta_\Est}{2}$ and $\Pr(\Err^x_\Est \ | \ \neg \Fail^x_\Est) \leq \frac{\delta_\Est}{2}$. Therefore, we can say that $$\Pr(\Err^x_\Est) \leq \Pr(\Fail^x_\Est) + \Pr(\Err^x_\Est \ | \ \neg \Fail^x_\Est) \leq \delta_\Est.$$
\end{proof}

Now we will show that for an arbitrary element $x \in [n]$, if its estimated probability mass $\pest{x}$ in the distribution $P$ is approximated well, then the ratios $\frac{\known(x)}{\unknown(x)}$ and $\frac{\known(x)}{\pest{x}}$ will be close to each other.

\begin{cl}\label{bclaim:estclosel1}
Let $\pest{x}$ be the estimate of $\unknown(x)$ such that $(1-\zeta)\unknown(x) \leq \pest{x} \leq (1+\zeta)\unknown(x)$ holds. Then $\E\limits_{x \sim P}\left[\size{\frac{\known(x)}{\unknown(x)} - \frac{\known(x)}{\pest{x}}}\right] \leq \frac{\zeta}{1 - \zeta}$.
\end{cl}
    
\begin{proof}
We know that $(1-\zeta)\unknown(x) \leq \pest{x} \leq (1+\zeta)\unknown(x)$. Hence we have
\begin{eqnarray*}
\frac{1}{\unknown(x)} - \frac{1}{(1-\zeta)\unknown(x)} 
& \leq & \frac{1}{\unknown(x)} -  \frac{1}{\pest{x}} \\
&\leq & \frac{1}{\unknown(x)} - \frac{1}{(1+\zeta)\unknown(x)} 
\end{eqnarray*}
    
Therefore 
\begin{eqnarray*}
&&\size{\frac{1}{\unknown(x)} -  \frac{1}{\pest{x}}} \leq  \frac{\zeta}{(1-\zeta) \unknown(x)}
\end{eqnarray*}
    
Therefore, by multiplying $\known(x)$ on both sides and taking expectation we have, 
\begin{eqnarray*}
\E\limits_{x \sim P}\left[\size{\frac{\known(x)}{\unknown(x)} -  \frac{\known(x)}{\pest{x}}}\right] \leq \frac{\zeta}{1-\zeta}\cdot \E\limits_{x \sim P}\left[\frac{\known(x)}{\unknown(x)}\right] = \frac{\zeta}{1-\zeta}.
\end{eqnarray*}
    
This completes the proof of the claim.
\end{proof}

\begin{cl}\label{lem:inftvbound}
If $\linf(\unknown, \known) \leq \eps_1$, and if $\Err^x_\Est$ does not occur for any sample $x \in \samples$ obtained in \infident, then with probability at least $1-\frac{\delta}{2}$, \infident computes $\dest$ such that $|\dest - \dtv(\unknown, \known)| \leq \frac{\eta - \eps}{2}$ holds.
\end{cl}

\begin{proof}
Following the promise of the claim, since $\Err^x_\Est$ does not occur for any sample $x \in \samples$, we know that \Est has not returned $\bot$ for any $x \in \samples$, where $\samples$ is the multi-set of samples obtained in \infident. Moreover, since $\linf(\unknown, \known) \leq \eps_1$, from the definition, we can say that for all $x \in [n]$, $(1-\eps_1)\unknown(x) \leq \known(x) \leq (1+\eps_1)\unknown(x)$ holds. Therefore, from \cref{lem:est_app} we have, $(1-\zeta)\unknown(x) \leq \pest{x} \leq (1+\zeta)\unknown(x)$, where $\pest{x}$ denotes the estimated probability of $x$ of the distribution $P$. Now let us divide the entire proof into several cases.

Let us first consider the case when $\known(x) < \unknown(x) < \pest{x}$. Then we have $\frac{\known(x)}{\pest{x}} < \frac{\known(x)}{\unknown(x)} < 1$. Therefore, using the triangle inequality, we can say the following: 
\begin{eqnarray*}
    \size{1 - \frac{\known(x)}{\pest{x}}} &\leq& \size{1 - \frac{\known(x)}{\unknown(x)}} + \size{\frac{\known(x)}{\unknown(x)} - \frac{\known(x)}{\pest{x}}} \\
    1 - \frac{\known(x)}{\pest{x}} &\leq& 1 - \frac{\known(x)}{\unknown(x)} + \size{\frac{\known(x)}{\unknown(x)} - \frac{\known(x)}{\pest{x}}} \\
    \mbox{So,}~~~~\max\left(0, 1 - \frac{\known(x)}{\pest{x}}\right) &\leq& \max\left(0, 1 - \frac{\known(x)}{\unknown(x)}\right) \\ \hspace{50pt} \ &+& \size{\frac{\known(x)}{\unknown(x)} - \frac{\known(x)}{\pest{x}}}
\end{eqnarray*}  

In a similar fashion, all the other 5 cases: $\known(x) < \pest{x} < \unknown(x)$, $\pest{x} < \unknown(x) < \known(x)$, $\unknown(x) < \known(x) < \pest{x}$, $\unknown(x) < \pest{x} < \known(x)$, $\pest{x} < \known(x) < \unknown(x)$ can be handled. Therefore, for all $x \in \samples$, we have:
\begin{eqnarray*}
\max\left(0, 1 - \frac{\known(x)}{\pest{x}}\right) &\leq& \max\left(0, 1 - \frac{\known(x)}{\unknown(x)}\right)  + \size{\frac{\known(x)}{\unknown(x)} - \frac{\known(x)}{\pest{x}}} 
\end{eqnarray*}

    Similarly, we also can show the following:
    \begin{eqnarray*}
        \max\left(0, 1 - \frac{\known(x)}{\pest{x}}\right) &\geq& \max\left(0, 1 - \frac{\known(x)}{\unknown(x)}\right)  - \size{\frac{\known(x)}{\unknown(x)} - \frac{\known(x)}{\pest{x}}} 
    \end{eqnarray*}

Therefore taking expectation with respect to $x$ sampled from $\unknown$ and using \Cref{bclaim:estclosel1}, we have the following inequality:
\begin{eqnarray}
\dtv(\unknown, \known) - \frac{\zeta}{1-\zeta} &\leq& \E\limits_{x \sim P}\left[\max\left(0, 1 - \frac{\known(x)}{\pest{x}}\right)\right] \leq \dtv(\unknown, \known) + \frac{\zeta}{1-\zeta} \label{eqn:linftvexpectationbound}   \\
\dtv(\unknown, \known) - \frac{\eta-\eps}{2} &\leq& \E\limits_{x \sim P}\left[\max\left(0, 1 - \frac{\known(x)}{\pest{x}}\right)\right] \leq \dtv(P, Q) + \frac{\eta-\eps}{2} 
\end{eqnarray}

The last inequality follows from $\zeta = \frac{\eta - \eps}{\eta - \eps + 2}$. Recall that $t =\frac{8}{ (\eta - \eps)^2} \log \frac{4}{\delta} $. Therefore, using the Chernoff-Hoeffding bound we can say that 
\begin{eqnarray*}
    \Pr\left(\dest > \dtv(P, Q) + \frac{\eta-\eps}{2} \right) 
    &\leq& \exp\left(-\left(\frac{\eta-\eps}{2}\right)^2 \cdot \frac{t^2 }{2t}\right) \\ 
    &\leq& \frac{\delta}{4}
\end{eqnarray*}

Similarly, we can prove the following:
$$\Pr \left(\dest < \dtv(P, Q) - \frac{\eta-\eps}{2}\right) \leq \frac{\delta}{4}$$

Combining the above two equations, we can conclude that with probability at least $1-\frac{\delta}{2}$, \infident computes $\dest$ such that $|\dest - \dtv(\unknown, \known)| \leq \frac{\eta - \eps}{2}$ holds.
\end{proof}

Now we are ready to prove the completeness lemma of \infident.

\begin{proof}[Proof of \cref{lem:completeness}]
Let us define the event $\Err_\Est$, which considers the failure of \Est for all sampled elements $x \in \samples$.
$$\Err_\Est := \cup_{x \in \cX} \Err_\Est^x$$
Since $\linf(\unknown, \known) \leq \eps_1$, therefore, $\dtv (\unknown, \known) \leq \eps$ holds. Thus using \cref{lem:inftvbound}, we can say that $\Pr\left(\dest > \frac{\eta + \eps}{2} \mid \neg \Err_\Est \right)$ is bounded by $\frac{\delta}{2}$. So, we have
\[\Pr\left(\mbox{\infident outputs }\reject\right) \leq \Pr\left(\dest > \frac{\eta + \eps}{2} \mid \neg \Err_\Est\right) + \Pr(\Err_\Est) \leq \frac{\delta}{2} + |\cX|\cdot\delta_\Est = \frac{\delta}{2} + t\cdot\delta_\Est \leq \delta.\]
The last inequality follows from the fact that $\delta_\Est = \frac{\delta}{4}$. This completes the proof of the lemma.
\end{proof} 

\subsubsection*{Soundness of \infident}
Let us now move towards proving the soundness of our algorithm. We will prove the following lemma.

\begin{lem}\label{lem:soundness}
Let $\dtv(\unknown, \known)\geq \eta$. Then with probability at least $1-\delta$, our algorithm \infident outputs \reject.
\end{lem}

\begin{proof}[Proof of \cref{lem:soundness}]
Let us first define an event 
$$\Early:= \Est \mbox{ early returns } \bot \mbox{ for any of the } x \in \cX.$$ 

From \Cref{lem:est_app}, we know that with probability $1-\frac{\delta_\Est}{2}$, $(1-\zeta)\unknown(x) \leq \pest{x} \leq (1+\zeta)\unknown(x)$ holds. Then since $\dtv(P,Q)\geq \eta$, similar to  \Cref{lem:inftvbound}, we can say that
\begin{align*}
\Pr\left(\dest < \frac{\eta+\eps}{2} \mid \neg \Early\right) \leq \frac{\delta}{2}
\end{align*}
    
Therefore, 
\begin{align*}
\Pr\left(\mbox{\infident outputs }\reject\right) &= \Pr\left(\dest > \frac{\eta+\eps}{2} \mid \neg \Early\right) \cdot \Pr(\neg \Early) + \Pr(\Early)\\
&\geq (1-\delta)\left(1-\Pr(\Early)\right) + \Pr(\Early)\\
&\geq 1 - \delta
\end{align*}
This completes the proof of the lemma.
\end{proof}

\subsubsection*{Query Complexity of \infident}

Note that in our algorithm \tpa, we call the algorithm \cintcond. \cintcond in turn calls the \intcond oracle to simulate samples.

\begin{lem}\label{lem:linftvquery}
Total number of \intcond queries performed by \infident is \[
    \tOh\left( \frac{1}{(\eta - \eps)^4} \E_{x \sim \unknown} \left[\tilt_\known(x) \cdot\min\left( \log^2 \frac{1}{\unknown(x)}, \log^2 \frac{1}{\known(x)} \right) \right] \right).
    \] in expectation.    
\end{lem}

\begin{proof}
Since $\theta = \Oh(\tilt_\known(x))$, therefore, using \cref{cl:estquery_app}, the expected number of \intcond queries made by \Est for a fixed $x$ is
\[\tOh\left(\frac{1}{(\eta - \eps)^4} \E_{x\sim \unknown}\left [ \tilt_\known(x) \cdot  \min\left(\log^2 \frac{1}{\unknown(x)}, \log^2 \frac{1}{\known(x)}\right)\right]\right).\]
\end{proof}

\begin{proof}[Proof of \cref{theo:infidentproof_app}]
    The proof follows from \cref{lem:completeness}, \cref{lem:soundness}, \cref{lem:linftvquery}. 
\end{proof}

\subsubsection{Proof of \cref{coro:tolerantuniformity}}

\tolerantuniformity*

To prove the above corollary, we will use the following lemma.

\begin{lem}
    $\E_{x\sim \unknown}\left[\log^2 \frac{1}{\unknown(x)}\right] \leq \log^2|S| + 2$ where $S$ is the support of $\unknown$.
    \label{lem:log2bound}
\end{lem}
\begin{proof}
    Number of $x$ such that $\unknown(x) \geq \frac{1}{e}$ is at most 2. Therefore, 
    \begin{align*}
        \E_{x\sim \unknown}\left[\log^2 \frac{1}{\unknown(x)}\right] &= \E_{x\sim \unknown | \unknown(x) \leq \frac{1}{e}}\left[\log^2 \frac{1}{\unknown(x)}\right] + \E_{x\sim \unknown | \unknown(x) > \frac{1}{e}}\left[\log^2 \frac{1}{\unknown(x)}\right]\\
        &\leq \log^2|S| + 2 \log^2 e = \log^2|S| +2
    \end{align*}
    Because $\log^2x$ is concave for $x>e$, by Jensen's inequality, 
     \begin{align*}
         \E_{x\sim \unknown | \unknown(x) \leq \frac{1}{e}}\left[\log^2 \frac{1}{\unknown(x)}\right] &= \sum_{x\sim \unknown | \unknown(x) \leq \frac{1}{e}} \unknown(x)\log^2 \frac{1}{\unknown(x)}\\
         &\leq \log^2\left(\sum_{x\sim \unknown | \unknown(x) \leq \frac{1}{e}} \unknown(x)\cdot \frac{1}{\unknown(x)}\right) \leq \log^2 |S|
     \end{align*}
\end{proof}

\begin{proof}[Proof of \cref{coro:tolerantuniformity}]
    Since $\tilt_U(x) = 1$ for all $x \in S$,and $\min\left(\log^2 \frac{1}{\unknown(x)}, \log^2 \frac{1}{U(x)}\right) \leq \log^2 \frac{1}{\unknown(x)}$ for all $x \in S$, the proof follows from \cref{theo:infidentproof_app} and \cref{lem:log2bound}.
\end{proof}
\section{Log Concave Distributions}

In this section, we study the impact of our results on the important class of log-concave distributions. Log-concave distributions are a widely studied class of structured distributions that generalize many common distributions, including uniform, binomial, geometric, and Poisson distributions. For concreteness, we consider the log-concave distributions to be defined over the infinite support $\Omega =\mathbb{Z}^+$. Any other support such as $\{0, \ldots, n\}$ or $\mathbb{Z}$ can be easily handled by appropriate shifting and truncation.

\begin{defi}[Log-Concave Distribution]
    A discrete distribution \(P\) over \(\Omega\) is said to be \emph{log-concave} if for every integer \(k \geq 1\), we have
    \[
    P(k)^2 \geq P(k-1) \cdot P(k+1).
    \]
\end{defi}

\begin{lem}
    \label{lem:simplified_tilt}
    For a log-concave distribution \(P\), $\tilt_P(x)$ can be simplified as follows:
    \[
    \tilt_P(x)=\max\!\Big(\frac{P(x-1)}{P(x)},\;\frac{P(x+1)}{P(x)}\Big).
    \]
    In other words, under log-concavity the maxima in the definition of \(\mathrm{tilt}_P(x)\)
    are attained at \(y=x-1\) and \(y=x+1\).
\end{lem}
\begin{proof}
    Recall that if \(P\) is log-concave, then for every integer $k \in [n]$, we have $P(k)^2 \geq P(k-1) \cdot P(k+1)$. This implies that the sequence $\frac{P(k-1)}{P(k)}$ is non-decreasing in $k$ and the sequence $\frac{P(k+1)}{P(k)}$ is non-increasing in $k$. Therefore for any $y \leq x - 1$, we have \[\frac{P(y)}{P([y+1, x])} \leq \frac{P(y)}{P(y+1)} \leq \frac{P(x-1)}{P(x)}.\] Similarly, for any $y \geq x + 1$, we have \[\frac{P(y)}{P([x, y-1])} \leq \frac{P(y)}{P(y-1)} \leq \frac{P(x+1)}{P(x)}.\] This completes the proof.
\end{proof}

\begin{lem}
    \label{lem:logconcave_tilt}
    Let $P$ be a Poisson distribution with parameter $\lambda >0$. Then, 
    \[
    \E_{x\sim P}\left[\tilt_P(x) \cdot \log^2\frac{1}{P(x)}\right] = \Oh\left(\log^2 \lambda\right).
    \]
\end{lem}

\begin{proof}
    Using Cauchy Schwarz inequality, we have
    \begin{align}
        \label{eq:cauchy_schwarz}
        \E_{x\sim P}\left[\tilt_P(x) \cdot \log^2\frac{1}{P(x)}\right] & \leq \sqrt{\E_{x\sim P}\left[\tilt_P^2(x)\right] \cdot \E_{x\sim P}\left[\log^4\frac{1}{P(x)}\right]}
    \end{align}
    Now let us first bound $\E_{x\sim P}\left[\tilt_P^2(x)\right]$. Using \Cref{lem:simplified_tilt}, we have that for any $x \in [n]$, $\tilt_P(x) = \max\left(\frac{P(x-1)}{P(x)}, \frac{P(x+1)}{P(x)}\right)$. For Poisson distribution with parameter $\lambda >0$, we have that 
    \begin{align*}
        \frac{P(x-1)}{P(x)} = \frac{x}{\lambda}, \ \ \text{and} \ \
        \frac{P(x+1)}{P(x)} = \frac{\lambda}{x+1}.
    \end{align*}
    Therefore for $x \geq \lambda$, we have $\tilt_P(x) = \frac{x}{\lambda}$, and for $x < \lambda$, we have $\tilt_P(x) = \frac{\lambda}{x+1}$ and for $x = 0$, we have $\tilt_P(0) = \lambda$. Thus, we have

    \begin{align*}
        \E_{x\sim P}\left[\tilt_P^2(x)\right] & = \sum_{x=0}^{\lfloor \lambda \rfloor -1} P(x) \cdot \left(\frac{\lambda}{x+1}\right)^2 + \sum_{x=\lceil \lambda \rceil}^{\infty} P(x) \cdot \left(\frac{x}{\lambda}\right)^2\\
        & = \lambda^2 \cdot \sum_{x=0}^{\lfloor \lambda \rfloor -1} P(x) \cdot \frac{1}{(x+1)^2} + \frac{1}{\lambda^2} \cdot \sum_{x=\lceil \lambda \rceil}^{\infty} P(x) \cdot x^2\\
        & \leq \lambda^2 \cdot \sum_{x=0}^{\infty} P(x) \cdot \frac{1}{(x+1)^2} + \frac{1}{\lambda^2} \cdot \sum_{x=0}^{\infty} P(x) \cdot x^2\\
        & = \lambda^2 \cdot \E_{x\sim P}\left[\frac{1}{(x+1)^2}\right] + \frac{1}{\lambda^2} (\lambda^2 + \lambda)
    \end{align*}
    Let us for now bound $\E_{x\sim P}\left[\frac{1}{(x+1)^2}\right]$. Let us split at $x+1 = \frac{\lambda}{2}$ and we have
    \begin{align*}
        \E_{x\sim P}\left[\frac{1}{(x+1)^2}\right] & = \sum_{x: x+1 \leq \frac{\lambda}{2}} P(x) \cdot \frac{1}{(x+1)^2} + \sum_{x: x+1 > \frac{\lambda}{2}} P(x) \cdot \frac{1}{(x+1)^2}
    \end{align*}
    For $x+1 > \frac{\lambda}{2}$, we have $\frac{1}{(x+1)^2} \leq \frac{4}{\lambda^2}$, therefore, 
    \[
    \E_{x\sim P}\left[\frac{1}{(x+1)^2}\right] \leq \sum_{x: x+1 \leq \frac{\lambda}{2}} P(x) \cdot \frac{1}{(x+1)^2} + \frac{4}{\lambda^2}.
    \]
    For $x+1 \leq \frac{\lambda}{2}$, we use the trivial bound of $\frac{1}{(x+1)^2} \leq 1$. Therefore, we have
    \[
    \E_{x\sim P}\left[\frac{1}{(x+1)^2}\right] \leq \mathbb{P}\left[x+1 \leq \frac{\lambda}{2}\right] + \frac{4}{\lambda^2}.
    \]
    Using Chernoff bound for Poisson distribution, we have that
    \[
    \mathbb{P}\left[x+1 \leq \frac{\lambda}{2}\right] \leq e^{-\lambda/8}.
    \]
    Hence, 
    \[
    \E_{x\sim P}\left[\frac{1}{(x+1)^2}\right] \leq e^{-\lambda/8} + \frac{4}{\lambda^2} = \Oh\left(\frac{1}{\lambda^2}\right).
    \]
    Therefore, we have
    \begin{align*}
        \E_{x\sim P}\left[\tilt_P^2(x)\right] & \leq \lambda^2 \cdot \Oh\left(\frac{1}{\lambda^2}\right) + \frac{1}{\lambda^2} (\lambda^2 + \lambda) = \Oh(1).
    \end{align*}
    In a similar way, by splitting the domain into two parts: $A = \{x : |x-\lambda| \leq \sqrt{C \lambda\log\lambda}\}$ and $\overline{A} = \{x : |x-\lambda| > \sqrt{C \lambda\log\lambda}\}$ for some $C>1$ we can show that $\E_{x\sim P}\left[\log^4\frac{1}{P(x)}\right] = \tOh(\log^4 \lambda)$. 
    Therefore, substituting the above results in \Cref{eq:cauchy_schwarz}
    \begin{align*}
        \E_{x\sim P}\left[\tilt_P(x) \cdot \log^2\frac{1}{P(x)}\right] & \leq \sqrt{\E_{x\sim P}\left[\tilt_P^2(x)\right] \cdot \E_{x\sim P}\left[\log^4\frac{1}{P(x)}\right]} = \tOh(\log^2 \lambda).
    \end{align*}
\end{proof}

\begin{restatable}{lem}{tolerantlogconcave}\label{coro:tolerantlogcon}  
Let $\unknown$ be an unknown distribution over $\mathbb{Z}$, and let $\known$ be Poisson distribution with parameter $\lambda >0$. Our algorithm $\infident$ can distinguish between $\linf(\unknown, \known)\leq 2\eps$ and $\dtv(\unknown,\known) \geq \eta$, with at most $\tOh\left(\frac{\log^2 \lambda}{(\eta-\eps)^4}\right)$ queries to the interval conditioning oracle in expectation.
\end{restatable}

\begin{proof}
    By observing that $\min\left(\tilt_\unknown(x), \theta\right) \leq \theta = O(\tilt_\known(x))$ for any $x \in \mathbb{Z}$, the proof follows directly from \Cref{thm:tvident} and \Cref{lem:logconcave_tilt}.
\end{proof}

\section{Details of Experiments}\label{sec:expresult_app} 

We start with recalling the implementation of \intcond and its proof.

\expcl*
\begin{proof}
    Since \samp{} draws $x = \chat^{-1}(u)$ for $u \sim \uni^{\contint{0,1}}$ and $\chat$ is monotonically increasing, the condition $x \in [a, b]$ is equivalent to $u \in \contint{\chat(a), \chat(b)}$. Hence, conditioning \samp{}'s output to $[a, b]$ is equivalent to sampling $u$ from $\uni^{\contint{\chat(a), \chat(b)}}$ and returning $\chat^{-1}(u)$.
\end{proof}

\subsection{Extended Experimental Results}

In this section, we present the case study to evaluate the effectiveness of \tester{} in detecting incorrect implementations of Binomial and Poisson samplers. We used the \infident{} mode in this setting. To simulate buggy implementations, we manually introduced errors by perturbing the constants used in their inverse sampling routines. For each distribution, we created six implementations: Implementation 1 (see \cref{fig:orig_sampler}) is correct, while Implementations 2–6 contain injected bugs. The buggy Binomial implementations correspond to \cref{fig:binbug_1} and \cref{fig:binbug_2}, and the buggy Poisson implementations correspond to \cref{fig:poibug_1} and \cref{fig:poibug_2}. For the implementation 5 and 6 (see \cref{fig:poibug_2}) of the Poisson sampler we apply benign changes by modifying the rejection sampling parameters that preserves correctness and should be accepted by our tester.

We analyze these buggy implementations of samplers, including hand-coded implementations. The results of these experiments are shown in \Cref{tab:binomial_exp} and \Cref{tab:poisson_exp}, where we report the outputs of our tool \tester{} as well as the number of sampler calls required by the tester. 
The Dec. column represents the outcome of \tester{}: `A' denoting an \accept{} and `R' denoting a \reject{}; the `\# Calls' column represents the number \intcond queries. The top half reports results for Binomial samplers, and the bottom half for Poisson samplers.
The plots clearly highlight how even minor deviations in implementation can lead to substantial statistical inaccuracies--detected by our tool.

\begin{figure}[h]
\centering
\begin{minipage}{0.49\linewidth}
\begin{lstlisting}[language=Python]
def Binomial(n, p):
    u = uniform(0, 1)
    if p == 0.0: return 0
    if p == 1.0: return n
    if n == 1: return int(u < p)

    spq = sqrt(n * p * (1 - p))
    b = 1.15 + 2.53 * spq
    a = -0.0873 + 0.0248 * b + 0.01 * p
    c = n * p + 0.5
    vr = 0.92 - 4.2 / b
    setup_complete = False

    while True:
        U = uniform(-0.5, 0.5)
        us = 0.5 - abs(U)
        k = floor((2 * a / us + b) * U + c)
        if k < 0 or k > n: continue
        V = uniform(0, 1)
        if us >= 0.07 and V <= vr: return k
        if not setup_complete:
            alpha = (2.83 + 5.1 / b) * spq
            lpq = log(p / (1 - p))
            m = floor((n + 1) * p)
            h = lgamma(m + 1) + lgamma(n - m + 1)
            setup_complete = True
        V *= alpha / (a / (us * us) + b)
        lk = lgamma(k + 1)
        lnk = lgamma(n - k + 1)
        if log(V) <= (h - lk - lnk + (k - m) * lpq):
            return k
\end{lstlisting}
\end{minipage}
\hfill
\begin{minipage}{0.49\linewidth}
\begin{lstlisting}[language=Python]
def Poisson(mu):
    if mu < 10:
        exlam = exp(mu)
        k = 0
        prod = 1
        while True:
            U = uniform(0,1)
            prod *= U
            if prod > exlam:
                k += 1
            else:
                return k
    else:
        lnlam = log(mu)
        b = 0.931 + 2.53 * sqrt(mu)
        a = -0.059 + 0.02483 * b
        vr = 0.9277 - 3.6224 / (b - 2)
        invalpha = 1.1239 + 1.1328 / (b - 3.4)
        
        while True:
            U = uniform(-0.5, 0.5)
            V = uniform(0, 1)
            lv = log(V)
            us = 0.5 - _fabs(U)
            k = floor((2 * a / us + b) * U + mu + 0.445)
            if (us >= 0.07) and (V <= vr): return k
            if (k <= 0) or (us < 0.013 and V > us): continue
            if (lv + log(invalpha) - log(a/us**2 + b)) <= k*lnlam - mu - lgamma(k):
                return k
\end{lstlisting}
\end{minipage}
\caption{\small Binomial and Poisson sampler implemented in GSL and NumPy.}
\label{fig:orig_sampler}
\end{figure}

\begin{figure}[t]
\centering
\begin{minipage}{0.49\linewidth}
\begin{lstlisting}[language=Python, escapechar=!]
def Binomial(n, p):
    if p == 0.0: return 0
    if p == 1.0: return n
    if n == 1: return int(uniform(-0.5, 0.5) < p)

    spq = sqrt(n * p * (1 - p))
    b = !\textcolor{red}{\underline {13.15}}! + 2.53 * spq
    a = -0.0874 + 0.0248 * b + 0.01 * p
    c = n * p + 0.5
    vr = 0.92 - 4.2 / b
    setup_done = False

    while True:
        U = uniform(-0.5, 0.5)
        us = 0.5 - abs(U)
        k = floor((2 * a / us + b) * U + c)
        if k < 0 or k > n: continue
        V = uniform(0, 1)
        if us >= 0.07 and V <= vr: return k
        if not setup_done:
            alpha = (2.83 + 5.1 / b) * spq
            lpq = log(p / (1 - p))
            m = floor((n + 1) * p)
            h = lgamma(m + 1) + lgamma(n - m + 1)
            setup_done = True
        V *= alpha / (a / (us * us) + b)
        lhs = log(V)
        rhs = h - lgamma(k + 1) - lgamma(n - k + 1) + (k - m) * lpq
        if lhs <= rhs: return k
\end{lstlisting}
\end{minipage} 
\hfill
\begin{minipage}{0.49\linewidth}
\begin{lstlisting}[language=Python, escapechar=!]
def Binomial(n, p):
    if p == 0.0: return 0
    if p == 1.0: return n
    if n == 1: return int(uniform(-0.5, 0.5) < p)

    spq = sqrt(n * p * (1 - p))
    b = !\textcolor{red}{\underline {13.15}}! + !\textcolor{red}{\underline {0.53}}! * spq
    a = -0.0874 + 0.0248 * b + 0.01 * p
    c = n * p + 0.5
    vr = 0.92 - 4.2 / b
    setup_done = False

    while True:
        U = uniform(-0.5, 0.5)
        us = 0.5 - abs(U)
        k = floor((2 * a / us + b) * U + c)
        if k < 0 or k > n: continue
        V = uniform(0, 1)
        if us >= 0.07 and V <= vr: return k
        if not setup_done:
            alpha = (2.83 + 5.1 / b) * spq
            lpq = log(p / (1 - p))
            m = floor((n + 1) * p)
            h = lgamma(m + 1) + lgamma(n - m + 1)
            setup_done = True
        V *= alpha / (a / (us * us) + b)
        lhs = log(V)
        rhs = h - lgamma(k + 1) - lgamma(n - k + 1) + (k - m) * lpq
        if lhs <= rhs: return k
\end{lstlisting}
\end{minipage}

\vspace{2ex}
\mbox{}  %

\begin{minipage}{0.49\linewidth}
\begin{lstlisting}[language=Python, escapechar=!]
def Binomial(n, p):
    if p == 0.0: return 0
    if p == 1.0: return n
    if n == 1: return int(uniform(-0.5, 0.5) < p)

    spq = sqrt(n * p * (1 - p))
    b = !\textcolor{red}{\underline {13.15}}! + !\textcolor{red}{\underline {0.53}}! * spq
    a = !\textcolor{red}{\underline {-0.1874}}! + 0.0248 * b + 0.01 * p
    c = n * p + 0.5
    vr = 0.92 - 4.2 / b
    setup_done = False

    while True:
        U = uniform(-0.5, 0.5)
        us = 0.5 - abs(U)
        k = floor((2 * a / us + b) * U + c)
        if k < 0 or k > n: continue
        V = uniform(0, 1)
        if us >= 0.07 and V <= vr: return k
        if not setup_done:
            alpha = (2.83 + 5.1 / b) * spq
            lpq = log(p / (1 - p))
            m = floor((n + 1) * p)
            h = lgamma(m + 1) + lgamma(n - m + 1)
            setup_done = True
        V *= alpha / (a / (us * us) + b)
        lhs = log(V)
        rhs = h - lgamma(k + 1) - lgamma(n - k + 1) + (k - m) * lpq
        if lhs <= rhs: return k
\end{lstlisting}
\end{minipage} 
\hfill
\begin{minipage}{0.49\linewidth}
\begin{lstlisting}[language=Python, escapechar=!]
def Binomial(n, p):
    if p == 0.0: return 0
    if p == 1.0: return n
    if n == 1: return int(uniform(-0.5, 0.5) < p)

    spq = sqrt(n * p * (1 - p))
    b = !\textcolor{red}{\underline {13.15}}! + !\textcolor{red}{\underline {0.53}}! * spq
    a = !\textcolor{red}{\underline {-0.0874}}! + !\textcolor{red}{\underline {0.148}}! * b + 0.01 * p
    c = n * p 
    vr = 0.92 - 4.2 / b
    setup_done = False

    while True:
        U = uniform(-0.5, 0.5)
        us = 0.5 - abs(U)
        k = floor((2 * a / us + b) * U + c)
        if k < 0 or k > n: continue
        V = uniform(0, 1)
        if us >= 0.07 and V <= vr: return k
        if not setup_done:
            alpha = (2.83 + 5.1 / b) * spq
            lpq = log(p / (1 - p))
            m = floor((n + 1) * p)
            h = lgamma(m + 1) + lgamma(n - m + 1)
            setup_done = True
        V *= alpha / (a / (us * us) + b)
        lhs = log(V)
        rhs = h - lgamma(k + 1) - lgamma(n - k + 1) + (k - m) * lpq
        if lhs <= rhs: return k
\end{lstlisting}
\end{minipage}
\caption{\small Flawed implementations of the Binomial sampler (Critical flaws are highlighted).}
\label{fig:binbug_1}
\end{figure}

\begin{figure}[t]
\centering
\begin{minipage}{0.49\linewidth}
\begin{lstlisting}[language=Python, escapechar=!]
def Binomial(n, p):
    if p == 0.0: return 0
    if p == 1.0: return n
    if n == 1: return int(uniform(-0.5, 0.5) < p)

    spq = sqrt(n * p * (1 - p))
    b = !\textcolor{red}{\underline {13.15}}! + !\textcolor{red}{\underline {0.53}}! * spq
    a = !\textcolor{red}{\underline {-0.0874}}! + !\textcolor{red}{\underline {0.148}}! * b + !\textcolor{red}{\underline {0.04}}! * p
    c = n * p + 0.5
    vr = 0.92 - 4.2 / b
    setup_done = False

    while True:
        U = uniform(-0.5, 0.5)
        us = 0.5 - abs(U)
        k = floor((2 * a / us + b) * U + c)
        if k < 0 or k > n: continue
        V = uniform(0, 1)
        if us >= 0.07 and V <= vr: return k
        if not setup_done:
            alpha = (2.83 + 5.1 / b) * spq
            lpq = log(p / (1 - p))
            m = floor((n + 1) * p)
            h = lgamma(m + 1) + lgamma(n - m + 1)
            setup_done = True
        V *= alpha / (a / (us * us) + b)
        lhs = log(V)
        rhs = h - lgamma(k + 1) - lgamma(n - k + 1) + (k - m) * lpq
        if lhs <= rhs: return k
\end{lstlisting}
\end{minipage}
\hfill
\hfill
\begin{minipage}{0.49\linewidth}
\begin{lstlisting}[language=Python, escapechar=!]
def Binomial(n, p):
    if p == 0.0: return 0
    if p == 1.0: return n
    if n == 1: return int(uniform(-0.5, 0.5) < p)

    spq = sqrt(n * p * (1 - p))
    b = 1.15 + !\textcolor{red}{\underline {0.53}}! * spq
    a = !\textcolor{red}{\underline {-0.0874}}! + !\textcolor{red}{\underline {0.148}}! * b + 0.01 * p
    !\textcolor{red}{\underline {c = n * p}}!
    vr = 0.92 - 4.2 / b
    setup_done = False

    while True:
        U = uniform(-0.5, 0.5)
        us = 0.5 - abs(U)
        k = floor((2 * a / us + b) * U + c)
        if k < 0 or k > n: continue
        V = uniform(0, 1)
        if us >= 0.07 and V <= vr: return k
        if not setup_done:
            alpha = (2.83 + 5.1 / b) * spq
            lpq = log(p / (1 - p))
            m = floor((n + 1) * p)
            h = lgamma(m + 1) + lgamma(n - m + 1)
            setup_done = True
        V *= alpha / (a / (us * us) + b)
        lhs = log(V)
        rhs = h - lgamma(k + 1) - lgamma(n - k + 1) + (k - m) * lpq
        if lhs <= rhs: return k
\end{lstlisting}
\end{minipage}
\caption{\small Flawed implementations of the Binomial sampler (Critical flaws are highlighted).}
\label{fig:binbug_2}
\end{figure}

\begin{figure}[h]
\centering
\begin{minipage}{0.49\linewidth}
\begin{lstlisting}[language=Python, escapechar=!]
def Poisson(mu):
    if mu < 10:
        exlam = exp(-mu)
        k = 0
        prod = 1
        while True:
            U = uniform(0,1)
            prod *= U
            if prod > exlam:
                k += 1
            else:
                return k
    else:
        lnlam = log(mu)
        b = !\textcolor{red}{\underline {1.931}}! + 2.53 * sqrt(mu)
        a = -0.059 + 0.02483 * b
        vr = 0.9277 - 3.6224 / (b - 2)
        invalpha = 1.1239 + 1.1328 / (b - 3.4)
        while True:
            U = uniform(-0.5, 0.5)
            V = uniform(0, 1)
            lv = log(V)
            us = 0.5 - fabs(U)
            k = floor((2 * a / us + b) * U + mu + 0.445)
            if (us >= 0.07) and (V <= vr): return k
            if (k <= 0) or (us < 0.013 and V > us): continue
            if (lv + log(invalpha) - log(a / us**2 + b)) <= k * lnlam - mu - lgamma(k):
                return k
\end{lstlisting}
\end{minipage}
\hfill
\hfill
\begin{minipage}{0.49\linewidth}
\begin{lstlisting}[language=Python, escapechar=!]
def Poisson(lambd):
    if lambd < 10:
        exlam = exp(-lambd)
        k = 0
        prod = 1
        while True:
            U = uniform(0,1)
            prod *= U
            if prod > exlam:
                k += 1
            else:
                return k
    else:
        lnlam = log(lambd)
        b = !\textcolor{red}{\underline {1.931 + 4.53}}! * sqrt(lambd)
        a = -0.059 + 0.02483 * b
        vr = 0.9277 - 3.6224 / (b - 2)
        invalpha = 1.1239 + 1.1328 / (b - 3.4)
        while True:
            U = uniform(-0.5, 0.5)
            V = uniform(0, 1)
            lv = log(V)
            us = 0.5 - fabs(U)
            k = floor((2 * a / us + b) * U + lambd + 0.445)
            if (us >= 0.07) and (V <= vr): return k
            if (k <= 0) or (us < 0.013 and V > us): continue
            if (lv + log(invalpha) - log(a / us**2 + b)) <= k * lnlam - lambd - lgamma(k):
                return k
\end{lstlisting}
\end{minipage}

\vspace{2ex}
\mbox{}  %

\begin{minipage}{0.49\linewidth}
\begin{lstlisting}[language=Python, escapechar=!]
def Poisson(lambd):
    if lambd < 10:
        exlam = exp(-lambd)
        k = 0
        prod = 1
        while True:
            U = uniform(0,1)
            prod *= U
            if prod > exlam:
                k += 1
            else:
                return k
    else:
        lnlam = log(lambd)
        b = !\textcolor{red} {\underline {1.931 + 4.53}}! * sqrt(lambd)
        a = !\textcolor{red} {\underline {-0.559}}! + 0.02483 * b
        vr = 0.9277 - 3.6224 / (b - 2)
        invalpha = 1.1239 + 1.1328 / (b - 3.4)

        while True:
            U = uniform(-0.5, 0.5)
            V = uniform(0, 1)
            lv = log(V)
            us = 0.5 - fabs(U)
            k = floor((2 * a / us + b) * U + lambd + 0.445)
            if (us >= 0.07) and (V <= vr): return k
            if (k <= 0) or (us < 0.013 and V > us): continue
            if (lv + log(invalpha) - log(a / us**2 + b)) <= k * lnlam - lambd - lgamma(k):
                return k
\end{lstlisting}
\end{minipage} 
\hfill
\begin{minipage}{0.49\linewidth}
\begin{lstlisting}[language=Python, escapechar=!]
def Poisson(lambd):
    if lambd < 10:
        exlam = exp(-lambd)
        k = 0
        prod = 1
        while True:
            U = uniform(0,1)
            prod *= U
            if prod > exlam:
                k += 1
            else:
                return k
    else:
        lnlam = log(lambd)
        b = !\textcolor{red} {\underline {1.931 + 4.53}}! * sqrt(lambd)
        a = !\textcolor{red} {\underline {-0.559 + 0.14483}}! * b
        vr = 0.9277 - 3.6224 / (b - 2)
        invalpha = 1.1239 + 1.1328 / (b - 3.4)

        while True:
            U = uniform(-0.5, 0.5)
            V = uniform(0, 1)
            lv = log(V)
            us = 0.5 - fabs(U)
            k = floor((2 * a / us + b) * U + lambd + 0.445)
            if (us >= 0.07) and (V <= vr): return k
            if (k <= 0) or (us < 0.013 and V > us): continue
            if (lv + log(invalpha) - log(a / us**2 + b)) <= k * lnlam - lambd - lgamma(k):
                return k
\end{lstlisting}
\end{minipage}
\caption{\small Flawed implementations of the Poisson sampler (Critical flaws are highlighted).}
\label{fig:poibug_1}
\end{figure}

\begin{figure}[t]
\centering
\begin{minipage}{0.49\linewidth}
\begin{lstlisting}[language=Python, escapechar=!]
def Poisson(lambd):
    if lambd < 10:
        exlam = exp(-lambd)
        k = 0
        prod = 1
        while True:
            U = uniform(0,1)
            prod *= U
            if prod > exlam:
                k += 1
            else:
                return k
    else:
        lnlam = log(lambd)
        b = 0.931 + 2.53 * sqrt(lambd)
        a = -0.059 + 0.02483 * b
        vr = 0.9277 - 3.6224 / (b - 2)
        invalpha = 1.1239 + 1.1328 / (b - 3.4)

        while True:
            U = uniform(-0.5, 0.5)
            V = uniform(0, 1)
            lv = log(V)
            us = 0.5 - fabs(U)
            k = floor((2 * a / us + b) * U + lambd + !\textcolor{red}{\underline {10.445}}!)
            if (us >= 0.07) and (V <= vr): return k
            if (k <= 0) or (us < 0.013 and V > us): continue
            if (lv + log(invalpha) - log(a / us**2 + b)) <= k * lnlam - lambd - lgamma(k):
                return k
\end{lstlisting}
\end{minipage} 
\hfill
\begin{minipage}{0.49\linewidth}
\begin{lstlisting}[language=Python, escapechar=!]
def Poisson(lambd):
    if lambd < 10:
        exlam = exp(-lambd)
        k = 0
        prod = 1
        while True:
            U = uniform(0,1)
            prod *= U
            if prod > exlam:
                k += 1
            else:
                return k
    else:
        lnlam = log(lambd)
        b = 0.931 + 2.53 * sqrt(lambd)
        a = -0.059 + 0.02483 * b
        vr = 0.9277 - 3.6224 / (b - 2)
        invalpha = !\textcolor{red} {\underline {100.1239}}! + 1.1328 / (b - 3.4)

        while True:
            U = uniform(-0.5, 0.5)
            V = uniform(0, 1)
            lv = log(V)
            us = 0.5 - fabs(U)
            k = floor((2 * a / us + b) * U + lambd + 0.445)
            if (us >= 0.07) and (V <= vr): return k
            if (k <= 0) or (us < 0.013 and V > us): continue
            if (lv + log(invalpha) - log(a / us**2 + b)) <= k * lnlam - lambd - lgamma(k):
                return k
\end{lstlisting}
\end{minipage}
\caption{\small Flawed implementations of the Poisson sampler (Critical flaws are highlighted).}
\label{fig:poibug_2}
\end{figure}

\begin{table}[h]
    \centering
\begin{tabular}{lcccccccccccc}
\toprule
(n, p) & \multicolumn{2}{c}{1} & \multicolumn{2}{c}{2} & \multicolumn{2}{c}{3} & \multicolumn{2}{c}{4} & \multicolumn{2}{c}{5} & \multicolumn{2}{c}{6} \\
 & Dec. & Calls & Dec. & Calls & Dec. & Calls & Dec. & Calls & Dec. & Calls & Dec. & Calls \\
\midrule
1000, 0.01 & \acceptcell & 31588 & \rejectcell & 38349 & \rejectcell & 6429122 & \rejectcell & 6042867 & \rejectcell & 6042726 & \rejectcell & 6042572 \\
3020, 0.02 & \rejectcell & 33391 & \acceptcell & 32137 & \rejectcell & 8799659 & \rejectcell & 20122177 & \rejectcell & 20122181 & \rejectcell & 20117736 \\
5040, 0.03 & \acceptcell & 35658 & \acceptcell & 37115 & \rejectcell & 9692837 & \rejectcell & 9902559 & \rejectcell & 9902559 & \rejectcell & 9912005 \\
7061, 0.04 & \acceptcell & 38292 & \rejectcell & 37776 & \rejectcell & 10085947 & \rejectcell & 10237109 & \rejectcell & 10237109 & \rejectcell & 10243369 \\
9081, 0.05 & \acceptcell & 37412 & \rejectcell & 34009 & \rejectcell & 10276355 & \rejectcell & 10397377 & \rejectcell & 10397377 & \rejectcell & 10397377 \\
11102, 0.06 & \acceptcell & 38726 & \rejectcell & 35105 & \rejectcell & 10366377 & \rejectcell & 10466423 & \rejectcell & 10466423 & \rejectcell & 10466423 \\
13122, 0.07 & \acceptcell & 39969 & \rejectcell & 34542 & \rejectcell & 10395893 & \rejectcell & 10480787 & \rejectcell & 10480787 & \rejectcell & 10483593 \\
15142, 0.08 & \acceptcell & 40765 & \rejectcell & 37233 & \rejectcell & 10388991 & \rejectcell & 10463841 & \rejectcell & 10463841 & \rejectcell & 10466163 \\
17163, 0.09 & \acceptcell & 41708 & \rejectcell & 37205 & \rejectcell & 10357631 & \rejectcell & 10426343 & \rejectcell & 10426343 & \rejectcell & 10426343 \\
19183, 0.1 & \acceptcell & 40261 & \rejectcell & 34950 & \rejectcell & 10308467 & \rejectcell & 10371085 & \rejectcell & 10371085 & \rejectcell & 10371085 \\
21204, 0.11 & \acceptcell & 42584 & \rejectcell & 37427 & \rejectcell & 22378565 & \rejectcell & 10304181 & \rejectcell & 10304181 & \rejectcell & 10305665 \\
23224, 0.12 & \acceptcell & 42628 & \rejectcell & 38402 & \rejectcell & 22355913 & \rejectcell & 10229975 & \rejectcell & 10229975 & \rejectcell & 10229975 \\
25244, 0.13 & \acceptcell & 43219 & \rejectcell & 37247 & \rejectcell & 10096557 & \rejectcell & 10146939 & \rejectcell & 10146939 & \rejectcell & 10148119 \\
27265, 0.14 & \acceptcell & 43140 & \rejectcell & 37539 & \rejectcell & 10012143 & \rejectcell & 10060781 & \rejectcell & 10060781 & \rejectcell & 10060781 \\
29285, 0.15 & \acceptcell & 44335 & \rejectcell & 39474 & \rejectcell & 9922735 & \rejectcell & 9968107 & \rejectcell & 9969077 & \rejectcell & 9969077 \\
31306, 0.16 & \acceptcell & 45536 & \rejectcell & 38395 & \rejectcell & 9829679 & \rejectcell & 9873785 & \rejectcell & 9873785 & \rejectcell & 9873785 \\
33326, 0.17 & \acceptcell & 44910 & \rejectcell & 38914 & \rejectcell & 9733201 & \rejectcell & 9775511 & \rejectcell & 9775511 & \rejectcell & 9775511 \\
35346, 0.18 & \acceptcell & 43482 & \rejectcell & 37734 & \rejectcell & 9634175 & \rejectcell & 9674359 & \rejectcell & 9674359 & \rejectcell & 9675119 \\
37367, 0.19 & \acceptcell & 44529 & \rejectcell & 39453 & \rejectcell & 9533053 & \rejectcell & 9571887 & \rejectcell & 9571887 & \rejectcell & 9571887 \\
39387, 0.2 & \acceptcell & 46805 & \rejectcell & 39645 & \rejectcell & 9429795 & \rejectcell & 9466869 & \rejectcell & 9466869 & \rejectcell & 9467527 \\
41408, 0.21 & \acceptcell & 47099 & \rejectcell & 39062 & \rejectcell & 9325027 & \rejectcell & 9360997 & \rejectcell & 9360997 & \rejectcell & 9361613 \\
43428, 0.22 & \acceptcell & 45767 & \rejectcell & 38580 & \rejectcell & 9218637 & \rejectcell & 9253689 & \rejectcell & 9253689 & \rejectcell & 9253689 \\
45448, 0.23 & \acceptcell & 45481 & \rejectcell & 38527 & \rejectcell & 12792894 & \rejectcell & 9144893 & \rejectcell & 9144893 & \rejectcell & 9144893 \\
47469, 0.24 & \acceptcell & 47470 & \rejectcell & 39353 & \rejectcell & 9179270 & \rejectcell & 9035117 & \rejectcell & 9035117 & \rejectcell & 9035631 \\
49489, 0.25 & \acceptcell & 48605 & \rejectcell & 40315 & \rejectcell & 8892519 & \rejectcell & 8924371 & \rejectcell & 8924857 & \rejectcell & 8924857 \\
51510, 0.26 & \acceptcell & 49311 & \rejectcell & 40627 & \rejectcell & 8782103 & \rejectcell & 8813329 & \rejectcell & 8813329 & \rejectcell & 8813791 \\
53530, 0.27 & \acceptcell & 46799 & \rejectcell & 39648 & \rejectcell & 8670761 & \rejectcell & 8701017 & \rejectcell & 8701017 & \rejectcell & 8701455 \\
55551, 0.28 & \acceptcell & 48555 & \rejectcell & 39208 & \rejectcell & 8559375 & \rejectcell & 8588533 & \rejectcell & 8588533 & \rejectcell & 8588533 \\
57571, 0.29 & \acceptcell & 47683 & \rejectcell & 40479 & \rejectcell & 8623971 & \rejectcell & 8474967 & \rejectcell & 8474967 & \rejectcell & 8475365 \\
59591, 0.3 & \acceptcell & 51027 & \rejectcell & 40400 & \rejectcell & 8332833 & \rejectcell & 8361061 & \rejectcell & 8361061 & \rejectcell & 8361061 \\
61612, 0.31 & \acceptcell & 48996 & \rejectcell & 40776 & \rejectcell & 8219231 & \rejectcell & 8246723 & \rejectcell & 8246723 & \rejectcell & 8246723 \\
63632, 0.32 & \acceptcell & 47735 & \rejectcell & 40061 & \rejectcell & 8105035 & \rejectcell & 8131871 & \rejectcell & 8131871 & \rejectcell & 8131871 \\
65653, 0.33 & \acceptcell & 49827 & \rejectcell & 41076 & \rejectcell & 7990397 & \rejectcell & 8016673 & \rejectcell & 8016673 & \rejectcell & 8016673 \\
67673, 0.34 & \acceptcell & 46644 & \rejectcell & 41355 & \rejectcell & 7875369 & \rejectcell & 7901035 & \rejectcell & 7901035 & \rejectcell & 7901035 \\
69693, 0.35 & \acceptcell & 48626 & \rejectcell & 40356 & \rejectcell & 7760029 & \rejectcell & 7784899 & \rejectcell & 7784899 & \rejectcell & 7785203 \\
71714, 0.36 & \acceptcell & 48515 & \rejectcell & 42042 & \rejectcell & 7822991 & \rejectcell & 7668815 & \rejectcell & 7668815 & \rejectcell & 7668815 \\
73734, 0.37 & \acceptcell & 51330 & \rejectcell & 42274 & \rejectcell & 7707066 & \rejectcell & 7552081 & \rejectcell & 7552081 & \rejectcell & 7552363 \\
75755, 0.38 & \acceptcell & 50089 & \rejectcell & 40401 & \rejectcell & 7412596 & \rejectcell & 7435425 & \rejectcell & 7435425 & \rejectcell & 7435425 \\
77775, 0.39 & \acceptcell & 50236 & \rejectcell & 41733 & \rejectcell & 7474391 & \rejectcell & 7318181 & \rejectcell & 7318181 & \rejectcell & 7318441 \\
79795, 0.4 & \acceptcell & 48312 & \rejectcell & 41421 & \rejectcell & 7178661 & \rejectcell & 7200841 & \rejectcell & 7200841 & \rejectcell & 7201093 \\
81816, 0.41 & \acceptcell & 49585 & \rejectcell & 39842 & \rejectcell & 7240701 & \rejectcell & 7083349 & \rejectcell & 7083593 & \rejectcell & 7083593 \\
83836, 0.42 & \acceptcell & 48904 & \rejectcell & 40957 & \rejectcell & 7123567 & \rejectcell & 6965827 & \rejectcell & 6965827 & \rejectcell & 6965827 \\
85857, 0.43 & \acceptcell & 50767 & \rejectcell & 41513 & \rejectcell & 6853953 & \rejectcell & 6847939 & \rejectcell & 6847939 & \rejectcell & 6847939 \\
87877, 0.44 & \acceptcell & 48425 & \rejectcell & 41447 & \rejectcell & 6709369 & \rejectcell & 6729813 & \rejectcell & 6729813 & \rejectcell & 6729813 \\
89897, 0.45 & \acceptcell & 50901 & \rejectcell & 41078 & \rejectcell & 6770837 & \rejectcell & 6611419 & \rejectcell & 6611631 & \rejectcell & 6611631 \\
91918, 0.46 & \acceptcell & 49892 & \rejectcell & 42053 & \rejectcell & 6653052 & \rejectcell & 6493147 & \rejectcell & 6493147 & \rejectcell & 6493351 \\
93938, 0.47 & \acceptcell & 50355 & \rejectcell & 42802 & \rejectcell & 6355579 & \rejectcell & 6374661 & \rejectcell & 6374661 & \rejectcell & 6374661 \\
95959, 0.48 & \acceptcell & 48603 & \rejectcell & 41369 & \rejectcell & 6237361 & \rejectcell & 6255905 & \rejectcell & 6255905 & \rejectcell & 6256097 \\
97979, 0.49 & \acceptcell & 49002 & \rejectcell & 40966 & \rejectcell & 6298416 & \rejectcell & 6137153 & \rejectcell & 6137153 & \rejectcell & 6137339 \\
100000, 0.5 & \acceptcell & 48472 & \rejectcell & 41695 & \rejectcell & 6180036 & \rejectcell & 6022699 & \rejectcell & 6019590 & \rejectcell & 9777344 \\
\bottomrule
\end{tabular}
    \caption{\small Run of \tester{} on the original Binomial sampler (1) and its buggy variants (2–6). The Dec. column represents the outcome of \tester{}: `A' denoting an \accept{} and `R' denoting a \reject{}; the `\# Calls' column represents the number \intcond queries.}
    \label{tab:binomial_exp}
\end{table}

\begin{table}[h]
    \centering
\begin{tabular}{lcccccccccccc}
\toprule
$\mu$ & \multicolumn{2}{c}{1} & \multicolumn{2}{c}{2} & \multicolumn{2}{c}{3} & \multicolumn{2}{c}{4} & \multicolumn{2}{c}{5} & \multicolumn{2}{c}{6} \\
 & Dec. & Calls & Dec. & Calls & Dec. & Calls & Dec. & Calls & Dec. & Calls & Dec. & Calls \\
\midrule
1000 & \acceptcell & 811562 & \rejectcell & 833237 & \rejectcell & 842044 & \rejectcell & 140369 & \rejectcell & 808156 & \acceptcell & 800560 \\
3020 & \acceptcell & 863672 & \rejectcell & 903475 & \rejectcell & 895914 & \rejectcell & 233807 & \acceptcell & 863148 & \acceptcell & 861109 \\
5040 & \acceptcell & 885827 & \rejectcell & 931999 & \rejectcell & 926929 & \rejectcell & 390081 & \acceptcell & 888461 & \acceptcell & 880820 \\
7061 & \acceptcell & 911888 & \rejectcell & 955740 & \rejectcell & 942692 & \rejectcell & 546425 & \acceptcell & 905637 & \acceptcell & 898524 \\
9081 & \acceptcell & 920237 & \rejectcell & 958480 & \rejectcell & 960679 & \rejectcell & 702723 & \acceptcell & 923127 & \acceptcell & 923894 \\
11102 & \acceptcell & 922181 & \rejectcell & 969748 & \rejectcell & 964924 & \rejectcell & 859039 & \acceptcell & 932612 & \acceptcell & 927106 \\
13122 & \acceptcell & 931155 & \rejectcell & 974499 & \rejectcell & 981279 & \rejectcell & 1015323 & \acceptcell & 935503 & \acceptcell & 941698 \\
15142 & \acceptcell & 943013 & \rejectcell & 991821 & \rejectcell & 988694 & \rejectcell & 1171595 & \acceptcell & 937303 & \acceptcell & 944408 \\
17163 & \acceptcell & 952475 & \rejectcell & 991288 & \rejectcell & 990015 & \rejectcell & 1327975 & \acceptcell & 952689 & \acceptcell & 945392 \\
19183 & \acceptcell & 954159 & \rejectcell & 1000092 & \rejectcell & 995948 & \rejectcell & 1484249 & \acceptcell & 962597 & \acceptcell & 953944 \\
21204 & \acceptcell & 960997 & \rejectcell & 995959 & \rejectcell & 1003561 & \rejectcell & 1640663 & \acceptcell & 964330 & \acceptcell & 959871 \\
23224 & \acceptcell & 968335 & \rejectcell & 998820 & \rejectcell & 1005720 & \rejectcell & 1796951 & \acceptcell & 963220 & \acceptcell & 966221 \\
25244 & \acceptcell & 970876 & \rejectcell & 1015614 & \rejectcell & 1011062 & \rejectcell & 1953229 & \acceptcell & 981683 & \acceptcell & 972422 \\
27265 & \acceptcell & 979746 & \rejectcell & 1018181 & \rejectcell & 1018462 & \rejectcell & 2109629 & \acceptcell & 984761 & \acceptcell & 969778 \\
29285 & \acceptcell & 981874 & \rejectcell & 1021841 & \rejectcell & 1022809 & \rejectcell & 2265891 & \acceptcell & 980396 & \acceptcell & 975532 \\
31306 & \acceptcell & 976914 & \rejectcell & 1022829 & \rejectcell & 1018280 & \rejectcell & 2422319 & \acceptcell & 982125 & \acceptcell & 977726 \\
33326 & \acceptcell & 992114 & \rejectcell & 1022337 & \rejectcell & 1018508 & \rejectcell & 2578567 & \acceptcell & 994920 & \acceptcell & 978024 \\
35346 & \acceptcell & 988439 & \rejectcell & 1033868 & \rejectcell & 1029464 & \rejectcell & 2734879 & \acceptcell & 993573 & \acceptcell & 990049 \\
37367 & \acceptcell & 988326 & \rejectcell & 1029277 & \rejectcell & 1038585 & \rejectcell & 2891319 & \acceptcell & 988753 & \acceptcell & 992263 \\
39387 & \acceptcell & 987044 & \rejectcell & 1039398 & \rejectcell & 1025142 & \rejectcell & 3047627 & \acceptcell & 1004585 & \acceptcell & 986729 \\
41408 & \acceptcell & 996146 & \rejectcell & 1038696 & \rejectcell & 1037226 & \rejectcell & 3204043 & \acceptcell & 999379 & \acceptcell & 995252 \\
43428 & \acceptcell & 1011172 & \rejectcell & 1044183 & \rejectcell & 1034100 & \rejectcell & 3360301 & \acceptcell & 1003038 & \acceptcell & 993015 \\
45448 & \acceptcell & 1007644 & \rejectcell & 1037111 & \rejectcell & 1046363 & \rejectcell & 3516679 & \acceptcell & 1005018 & \acceptcell & 997451 \\
47469 & \acceptcell & 1005712 & \rejectcell & 1052681 & \rejectcell & 1059253 & \rejectcell & 3673031 & \acceptcell & 1011685 & \acceptcell & 999759 \\
49489 & \acceptcell & 1000336 & \rejectcell & 1046995 & \rejectcell & 1045024 & \rejectcell & 3829305 & \acceptcell & 1008995 & \acceptcell & 1001455 \\
51510 & \acceptcell & 1009578 & \rejectcell & 1046716 & \rejectcell & 1056458 & \rejectcell & 3985759 & \acceptcell & 1009112 & \acceptcell & 995413 \\
53530 & \acceptcell & 1008156 & \rejectcell & 1048786 & \rejectcell & 1046096 & \rejectcell & 4142009 & \acceptcell & 1012053 & \acceptcell & 1007457 \\
55551 & \acceptcell & 1010670 & \rejectcell & 1050900 & \rejectcell & 1054924 & \rejectcell & 4298431 & \acceptcell & 1009956 & \acceptcell & 1003210 \\
57571 & \acceptcell & 1013733 & \rejectcell & 1056209 & \rejectcell & 1053975 & \rejectcell & 4454751 & \acceptcell & 1018643 & \acceptcell & 1014457 \\
59591 & \acceptcell & 1016898 & \rejectcell & 1056387 & \rejectcell & 1054065 & \rejectcell & 4611163 & \acceptcell & 1009524 & \acceptcell & 1008797 \\
61612 & \acceptcell & 1021055 & \rejectcell & 1051312 & \rejectcell & 1063780 & \rejectcell & 4767489 & \acceptcell & 1019301 & \acceptcell & 1020966 \\
63632 & \acceptcell & 1023954 & \rejectcell & 1059611 & \rejectcell & 1060530 & \rejectcell & 4923825 & \acceptcell & 1016645 & \acceptcell & 1009032 \\
65653 & \acceptcell & 1024920 & \rejectcell & 1069193 & \rejectcell & 1060997 & \rejectcell & 5080317 & \acceptcell & 1019177 & \acceptcell & 1021991 \\
67673 & \acceptcell & 1025479 & \rejectcell & 1070824 & \rejectcell & 1067935 & \rejectcell & 5236555 & \acceptcell & 1022327 & \acceptcell & 1016073 \\
69693 & \acceptcell & 1024368 & \rejectcell & 1067478 & \rejectcell & 1067391 & \rejectcell & 5392867 & \acceptcell & 1024166 & \acceptcell & 1016161 \\
71714 & \acceptcell & 1026069 & \rejectcell & 1070029 & \rejectcell & 1065405 & \rejectcell & 5549329 & \acceptcell & 1021577 & \acceptcell & 1030077 \\
73734 & \acceptcell & 1026201 & \rejectcell & 1075555 & \rejectcell & 1064427 & \rejectcell & 5705655 & \acceptcell & 1023129 & \acceptcell & 1013020 \\
75755 & \acceptcell & 1024594 & \rejectcell & 1067087 & \rejectcell & 1067052 & \rejectcell & 5862125 & \acceptcell & 1029113 & \acceptcell & 1022120 \\
77775 & \acceptcell & 1031796 & \rejectcell & 1073444 & \rejectcell & 1073015 & \rejectcell & 6018451 & \acceptcell & 1032416 & \acceptcell & 1028679 \\
79795 & \acceptcell & 1031057 & \rejectcell & 1069023 & \rejectcell & 1080605 & \rejectcell & 6174701 & \acceptcell & 1025333 & \acceptcell & 1025610 \\
81816 & \acceptcell & 1030519 & \rejectcell & 1076334 & \rejectcell & 1078865 & \rejectcell & 6331235 & \acceptcell & 1036999 & \acceptcell & 1034852 \\
83836 & \acceptcell & 1031874 & \rejectcell & 1074271 & \rejectcell & 1074054 & \rejectcell & 6487467 & \acceptcell & 1036302 & \acceptcell & 1037340 \\
85857 & \acceptcell & 1030004 & \rejectcell & 1083930 & \rejectcell & 1075691 & \rejectcell & 6643983 & \acceptcell & 1032076 & \acceptcell & 1030643 \\
87877 & \acceptcell & 1035003 & \rejectcell & 1075950 & \rejectcell & 1083257 & \rejectcell & 6800337 & \acceptcell & 1036899 & \acceptcell & 1032992 \\
89897 & \acceptcell & 1045326 & \rejectcell & 1077846 & \rejectcell & 1079941 & \rejectcell & 6956599 & \acceptcell & 1043417 & \acceptcell & 1030474 \\
91918 & \acceptcell & 1050357 & \rejectcell & 1076906 & \rejectcell & 1080035 & \rejectcell & 7112993 & \acceptcell & 1042812 & \acceptcell & 1035568 \\
93938 & \acceptcell & 1040430 & \rejectcell & 1079177 & \rejectcell & 1082469 & \rejectcell & 7269369 & \acceptcell & 1046877 & \acceptcell & 1041210 \\
95959 & \acceptcell & 1043863 & \rejectcell & 1082651 & \rejectcell & 1079669 & \rejectcell & 7425719 & \acceptcell & 1040584 & \acceptcell & 1043194 \\
97979 & \acceptcell & 1043282 & \rejectcell & 1080548 & \rejectcell & 1080236 & \rejectcell & 7582207 & \acceptcell & 1035713 & \acceptcell & 1045610 \\
100000 & \acceptcell & 1040765 & \rejectcell & 1080798 & \rejectcell & 1083447 & \rejectcell & 7738505 & \acceptcell & 1041750 & \acceptcell & 1031646 \\
\bottomrule
\end{tabular}
    \caption{\small Run of \tester{} on the original Poisson sampler (1) and its buggy variants (2–6). The Dec. column represents the outcome of \tester{}: `A' denoting an \accept{} and `R' denoting a \reject{}; the `\# Calls' column represents the number \intcond queries.}
    \label{tab:poisson_exp}
\end{table}

\end{document}